\documentclass[11pt]{article}
\usepackage{graphicx} 
\usepackage{amsmath, amsfonts,amsthm}
\usepackage[margin=1in]{geometry}
\usepackage{caption}
\usepackage{subcaption}
\usepackage{accents}
\usepackage{placeins}

\usepackage[hidelinks]{hyperref}
\hypersetup{
    colorlinks=true,
    linkcolor=blue,
    citecolor=blue
}

\usepackage[
backend=biber,style=apa,citestyle=authoryear,sortcites,sorting=nyt,date=year,labeldate=year,
maxbibnames=10,minbibnames=1,uniquename=false,uniquelist=false,maxcitenames=3]{biblatex}
\newtheorem{theorem}{Theorem}

\theoremstyle{plain}

\newtheorem{assumption}{Assumption}

\newtheorem{corollary}{Corollary}

\newtheorem{lemma}{Lemma}

\usepackage{float} 
\usepackage{csquotes}

\theoremstyle{definition}
\newtheorem{algorithm}{Algorithm}
\newtheorem{definition}{Definition}

\theoremstyle{remark}
\newtheorem{remark}{Remark}
\newtheorem{example}{Example}

\DeclareMathOperator{\interior}{0}

\title {Matching under Imperfectly Transferable Utility\thanks{We are grateful to the editors, to an anonymous reviewer and to Antoine Jacquet for useful comments.} \\[1ex] \large Submitted as a draft chapter for the \emph{Handbook of the Economics of Matching}, edited by Che, Chiappori and Salanié}
\author{Alfred Galichon and Simon Weber}
\date{\today}

\begin{document}
\maketitle

\section{Introduction}

\subsection{Motivating remarks}

In this chapter, we examine matching models with imperfectly transferable utility\footnote{We use the term ``imperfectly transferable utility'' throughout this chapter, but others have used terms such as matching with ``nontransferabilities'' or ``not fully transferable'' utility, see e.g. \textcite{legrosnewman2007}.}. We argue that imperfectly transferable utility is a useful generalization of transferable utility for modeling a number of matching markets for at least two reasons. First, in some settings, there are clear restrictions on the transfer technology that prevent utility from being perfectly transferable. For example, on the labor market, the wage paid by the firm differs from the wage received by the worker due to the (possibly nonlinear) tax schedule in place on that market. Second, the assumption of perfectly transferable utility is restrictive, despite its advantages. For example, suppose that on the marriage market, married men and women bargain over utility allocations in typical collective fashion. If utility is perfectly transferable, the model yields the same demand for public goods (e.g., expenditures on children) irrespective of the distribution of resources or allocation of bargaining power within the household.

Throughout the chapter, we will explore models outside of the transferable utility realm. By ``imperfectly transferable utility'' (ITU) we mean a situation where two potentially matched partners $i$ and $j$ bargain over a set of feasible utilities $(u_i,v_j)\in \mathcal F_{ij}$. The (perfectly) ``transferable utility'' (TU) case corresponds to $\mathcal F_{ij}=\{ (u_i,v_j): u_i+v_j \leq \Phi_{ij}\}$; but more generally, in the ITU case, the feasible set of utilities $\mathcal F_{ij}$ is defined by $\mathcal F_{ij}=\{ (u_i,v_j): A_{ij}(u_i)+B_{ij}(v_j) \leq \Phi_{ij} \}$ for some functions $A_{ij}$ and $B_{ij}$ which are increasing and continuous, but not necessarily linear.  

In some cases, the distinction between TU and ITU is quite innocuous. In particular, if $i$ and $j$ bargain in autarky with reservation utilities $\bar{u}$ and $\bar{v}$, then one can perform a rescaling of the utilities of both partners $\tilde{u}_i = A_{ij}(u_i)$ and $\tilde{v}_j = B_{ij}(v_j)$, express the reservation utilities in the new scale, and the problem becomes a TU one. In stark contrast, when instead of bargaining in isolation, partners are part of a matching market, things change drastically and this distinction becomes highly relevant. Assume that $i$ is in discussion with two potential partners $j_1$ and $j_2$; in that case, it will generally be impossible to find a new utility scale under which the bargaining problems with both $j_1$ and $j_2$ are TU. To quote \textcite{chiappori2017} (with our own emphasis), the TU setting ``allows the transfer of utility between agents at a constant ``exchange rate,'' so that, for a well-chosen cardinalization of individual preferences, increasing my partner's utility by one ``utile'' (i.e., unit of utility) has a cost of exactly one utile for me, \emph{irrespective of the economic environment} (...)''. In the matching setting studied in this book, the partner's choice is endogenous, and thus the cardinalization of individual preferences is no longer irrespective of the economic environment. 

In the remainder of this introduction, we present the notations and then discuss a few motivating examples. Throughout the chapter, the presentation is biased towards our own work. Our presentation of aggregate equilibrium in ITU models, as well as a number of related results and applications follows in large part the exposition in~\textcite{galichonkominersweber2019}, hereafter abbreviated GKW; we have also used material from~\textcite{chenchoogalichonetal2023} and~\textcite{chenchoogalichonetal2023b}

\subsection{Setting and notations}

We consider a general two-sided matching model with one-to-one assignment. To introduce our notation, we use the example of men and women\footnote{For the purposes of this chapter, we focus on the heterosexual marriage market. The homosexual marriage has been studied in e.g. \textcite{ciscatogalichongousse2020}.} matching on the marriage market. However, our notation can be adapted to other matching contexts, such as matching workers to jobs or children to childcare center positions\footnote{This chapter primarily focuses on one-to-one matching, though some of these examples are more representative of one-to-many scenarios in practice. \textcite{corblet2023} provides an extension of the \textcite{choosiow2006} framework to one-to-many matching in the TU case ; section \ref{sec:onetomany} in this chapter deals with one-to-many matching in the ITU case.
}.

In a  heterosexual marriage market, there are two populations of men and women who meet and may form pairs. Men are indexed by $i\in \mathcal{I}$ and women are indexed by $j\in \mathcal{J}$. At this individual level, a matching is a variable $\mu _{ij}$ equal to 1 if man $i$ and woman $j$ are matched, and 0 otherwise. Note that the matching must satisfy basic feasibility conditions, such as the requirement that each individual can only be matched to at most one partner, i.e. $\mu _{ij}\in \{ 0,1\}$, $\sum_{j\in \mathcal{J}}\mu _{ij}\leq 1$ and $\sum_{i\in \mathcal{I}}\mu _{ij}\leq 1$.

In a more ``macroscopic'' perspective, we may assume that men and women can be gathered in a finite number of types, which are defined as groups sharing
similar observable characteristics. We let $x_i\in \mathcal{X}$ and $y_j\in \mathcal{Y}$ denote the types of man $i$ and woman $j$, respectively. The number of types of men (resp. women) is $|\mathcal{X}|$ (resp. $|\mathcal{Y}|$). We let $n_x$ and $m_y$ denote the total mass of men of
type $x$ and women of type $y$, respectively. We introduce the sets $\mathcal{X}_{0}=\mathcal{X}%
\cup \mathcal{\{}0\mathcal{\}}$ and $\mathcal{Y}_{0}=\mathcal{Y}\cup
\mathcal{\{}0\mathcal{\}},$ where $0$ denotes the option of singlehood. The set of couple types is denoted $\mathcal{XY}=\mathcal{X\times Y},$ and the set of household types is denoted $\mathcal{XY}_{0}=\mathcal{X\times Y}\cup
\mathcal{X\times }\left\{ 0\right\} \cup \left\{ 0\right\} \times \mathcal{Y}
$. 

At the macroscopic level, an \emph{aggregate matching} is a vector $(\mu _{xy})_{x\in \mathcal{X},y\in
\mathcal{Y}}$ measuring the mass of matches between men of type $x$ and
women of type $y$. We let $\mathcal{M}$ denote the set of feasible matchings, that is, the
set of vectors $\mu _{xy}\geq 0$ such that $\sum_{y\in \mathcal{Y}}\mu _{xy}\leq
n_{x}$ and $\sum_{x\in \mathcal{X}}\mu _{xy}\leq m_{y}$. The strict interior of $\mathcal{M}$, denoted $%
\mathcal{M}^{\interior}$, is the set of matchings $\mu _{xy}>0$ such that $%
\sum_{y\in \mathcal{Y}}\mu _{xy}<n_{x}$ and $\sum_{x\in \mathcal{X}}\mu
_{xy}<m_{y}$. We call the elements of $\mathcal{M}^{\interior}$ \emph{interior matchings}. Finally, we let $\mu _{x0}$ and $\mu _{0y}$ denote the mass of single men of type $x$ and the mass
of single women of type $y$, respectively.

\subsection{A bestiary of models}
We motivate the use of ITU matching models through the following examples. Examples \ref{ex:introsimpleLTU} and \ref{ex:introintersectionLTU} examine a simple matching model of the labour market in which transfers (wages) are taxed at a single tax rate, or according to some convex tax schedule. Examples \ref{ex:introsimpleETU} and \ref{ex:introunionETU} introduce a simple matching model of the marriage market in which partners spend their income on private and public consumption.

\begin{example}[Matching with flat taxes]\label{ex:introsimpleLTU}

Consider a matching model in which workers (indexed by $i\in\mathcal{I}$) match with firms (indexed by $j\in\mathcal{J}$). Let $\alpha_{ij}$ be the amenities enjoyed by worker $i$ at firm $j$ and $\gamma_{ij}$ be the productivity of worker $i$ when working in firm $j$. We denote $w_{ij}$ the gross wage paid by the firm to the worker. Suppose wages are taxed at a rate of $\tau_{ij}$, then worker $i$ and firm $j$ receive utilities $u_i = \alpha_{ij} + (1-\tau_{ij})w_{ij}$ and $v_j = \gamma_{ij}-w_{ij}$, respectively. Let $\lambda_{ij} = 1/(1-\tau_{ij})$ and $\Phi_{ij}\equiv \lambda_{ij}\alpha_{ij}+\gamma_{ij}$. It is easy to show that the Pareto frontier, which describes the set of Pareto efficient utility allocations that worker $i$ and firm $j$ can achieve, is characterized by the following equation:
\begin{equation*}
    v_j = \Phi_{ij} - \lambda_{ij}u_i 
\end{equation*}
The frontier is a straight line, but not necessarily of slope $-1$ as in the TU case. Unless $\lambda_{ij} = \lambda$ for all $i,j$ pairs, it is not possible to find a cardinalization of the utilities such that frontier is of slope $-1$ for all $i$ and $j$. Later, we will refer to this model as the \textit{linearly transferable utility} (LTU) model.
\end{example}

\begin{example}[Matching with progressive taxes]\label{ex:introintersectionLTU}
Consider the same setting as in example \ref{ex:introsimpleLTU}, but this time assume that there are $K$ tax thresholds $t^1,...,t^K$ and $K+1$ tax rates. Any income between thresholds $t^{k}$ and $t^{k+1}$ is
taxed at rate $\tau ^{k}$, income above $t^{K}$ is taxed at rate $\tau
^{K}$, and income below $t^1$ is not taxed. We assume the tax schedule is convex, i.e. that the tax rates are increasing. It can be shown that the Pareto frontier is described by the equation
\begin{equation*}
    v_j = \min_{k\in\{0,...,K\}}\{\gamma_{ij} + \lambda^k\alpha_{ij}^k - \lambda^ku_i\}
\end{equation*}
where $\lambda^k = 1/(1-\tau^k)$ and where $\alpha_{ij}^k$ is a function of the amenities, the tax thresholds and tax rates. The frontier of the bargaining set is piecewise linear, and the bargaining set itself can be thought of as the intersection of many elementary bargaining sets of the type described in example \ref{ex:introsimpleLTU}.
\end{example}
\begin{example}[Matching with private consumption and marital affinity]\label{ex:introsimpleETU}

Consider a matching model of the marriage market in which men (indexed by $i\in\mathcal{I}$) match with women (indexed by $j\in\mathcal{J}$). Suppose that when a man $i$ and a woman $j$ form a match, they must decide on how to allocate their income $B_{ij}$ between private consumption for the man ($c_i$) and private consumption for the woman ($c_j$). For simplicity, we assume that men and women receive utilities $u_i = \alpha_{ij} + \tau_{ij}\log(c_i)$ and $v_j = \gamma_{ij} + \tau_{ij}\log(c_j)$, respectively, 
where the terms $\alpha$ and $\gamma$ capture marital affinities, $\tau$ is a preference parameter, and $c_i$ and $c_j$ are such that $c_i+c_j = B_{ij}$. In this model, the Pareto frontier is given by
\begin{equation*}
    v_j = \gamma_{ij} + \tau_{ij}\log\left(B_{ij} - \exp\left(\frac{u_i - \gamma_{ij}}{\tau_{ij}}\right)\right)
\end{equation*}
The frontier of the bargaining set is non-linear. Except in trivial cases in which $\alpha$, $\gamma$ and $\tau$ do not depend on $(i,j)$, it is not possible find a cardinalization of the utilities such that the frontier is a straight line of slope $-1$ for all $i$ and $j$. Later, we will refer to this model as the \textit{exponentially transferable utility} (ETU) model.
\end{example}

\begin{example}[Matching with private consumption, marital affinities and public goods]
\label{ex:introunionETU}
Consider the same setting as in example \ref{ex:introsimpleETU}, but assume that couples can decide on a discrete amount of public good $g_k\in\mathcal{G}$ and allocate the remaining income $B_{ij}(g_k)$ between  private consumption for the man and private consumption for the woman. Let the utilities received by men and women be $u_i = \alpha_{ij}(g_k) + \tau_{ij}\log(c_i)$ and $v_j = \gamma_{ij}(g_k) + \tau_{ij}\log(c_j)$, respectively. The equation of the Pareto frontier becomes
\begin{equation*}
    v_j = \max_{g_k\in\mathcal{G}}\left\{\gamma_{ij}(g_k) + \tau_{ij}\log\left(B_{ij}(g_k) - \exp\left(\frac{u_i - \gamma_{ij}(g_k)}{\tau_{ij}}\right)\right)\right\}
\end{equation*}
The frontier is once again non-linear. The bargaining set is typically not convex and can be thought of as the union of many elementary bargaining sets of the type described in example \ref{ex:introsimpleETU}.
\end{example}

These are just a few examples, but ITU models abound in the literature, for instance in the case of matching with investments (see the chapter of \textcite{noeldeke_samuelson_handbook} in this Handbook), models of family economics (see the chapter of \textcite{chiappori_low_handbook}), and matching with contracts (see the chapter of \textcite{hatfield_kominers_handbook}).

\section{Pairwise bargaining sets}

We begin by describing the utility allocations that a matched pair can agree on. We assume that when a man and a woman match, they decide on an outcome within a feasible (or bargaining) set of utility allocations. We describe the structure of these bargaining sets, for which we provide an implicit representation and illustrating our approach with several examples.

\subsection{The Pareto efficiency approach vs. the Pareto weights approach}

\begin{figure}[t]
\caption{Choice of public goods and randomization}
\centering
\includegraphics[scale=0.9]{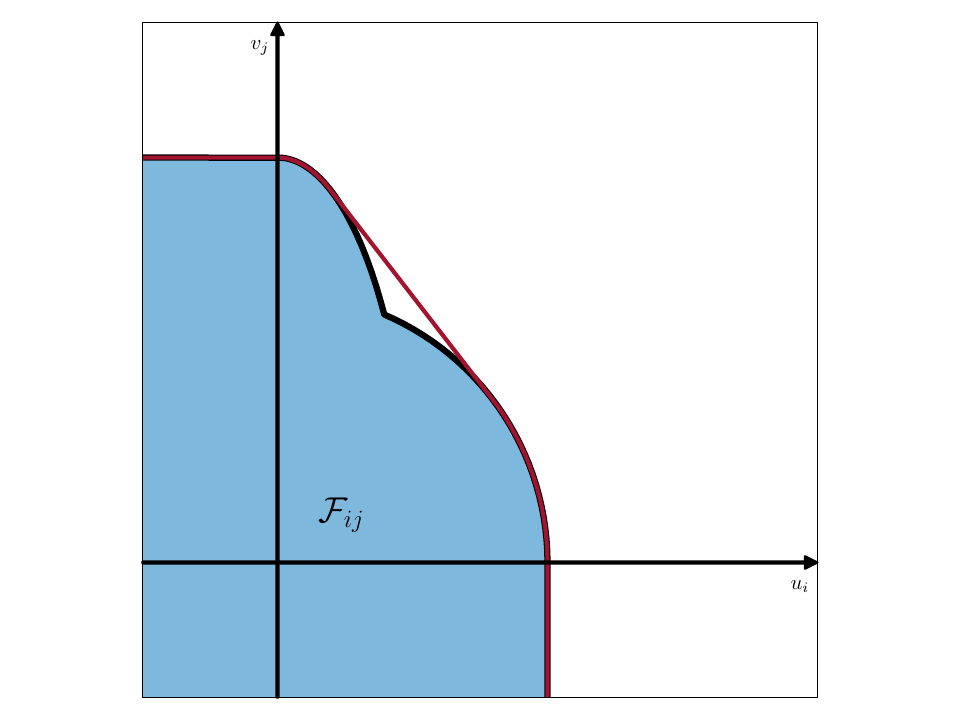}
\label{fig:ctu}
\end{figure}

The collective approach to household decision fundamentally relies on the
\textquotedblleft \textbf{Pareto efficiency  assumption}\textquotedblright ,
which states that couples make decisions that are Pareto efficient (see \textcite{delbocaflinn2012} or \textcite{browningchiapporiweiss2014} for discussions on the Pareto efficiency assumption\footnote{See  \textcite{prescotttownsend1984} for an early reference on Pareto efficiency and Pareto weights.}). In this
chapter's terminology, if $\mathcal{F}_{ij}$\ is the set of utilities $\left(
u_{i},v_{j}\right) $ that are jointly achievable by $i$ and $j$, then $i$
and $j$ will never pick a pair of utilities $\left( u_{i},v_{j}\right) $ which
is Pareto dominated. On figure~\ref{fig:ctu}, that means they lie on the Pareto 
frontier, which is the solid black line that envelopes the blue set $\mathcal F_{ij}$. 

While this assumption is arguably plausible, it is often replaced by a stronger
\textquotedblleft \textbf{Pareto weights assumption}\textquotedblright\
which states that couples pick pairs of utilities $\left( u_{i},v_{j}\right) 
$ which maximize the weighted sum of the utilities $\omega _{ij}u_{i}+\left(
1-\omega _{ij}\right) v_{j}$ over the set $\mathcal{F}_{ij}$. The weights $%
\omega _{ij}$ and $\left( 1-\omega _{ij}\right) $ are deemed the Pareto weights associated with $i$ and $j$ respectively. Geometrically, these points 
are obtained by taking the frontier of the convex hull of the feasible set; they 
are represented in the solid red line on figure~\ref{fig:ctu}. If the feasible set is
described by $\mathcal{F}_{ij}=\left\{ \left( u_{i},v_{j}\right) :v_{j}\leq
F_{ij}\left( u_{i}\right) \right\} $ where $F_{ij}\left( .\right) $ is
concave, the chosen utilities will be $\left( u_{i},v_{j}=F_{ij}\left(
u_{i}\right) \right) $ such that 
\[
\frac{\omega _{ij}}{1-\omega _{ij}}=-F_{ij}^{\prime }\left( u_{i}\right) ,
\]%
which obtains $u_{i}$ as a increasing function of $i$'s Pareto weight.

While the two approaches coincide in the case when $\mathcal{F}_{ij}$ is
convex, they are in
general not equivalent, as it is apparent on the figure. There as $\mathcal{F}_{ij}$ is not convex,
there are points on the Pareto frontier of $\mathcal{F}_{ij}$ which are
not on the frontier of the convex envelope $conv\left( \mathcal{F}%
_{ij}\right) $. These points will never be picked by the Pareto weights approach. 
But far from being a mathematical subtlety,
this can be a serious modeling issue, as we argue in the following example.

\FloatBarrier

Consider the problem of home ownership decision. A couple needs to 
decide whether to rent or buy their home. Assume that the feasible set of
partners' utilities $\left( u_{i},v_{j}\right) $ is $\mathcal{F}_{ij}^{R}$
if they rent, and $\mathcal{F}_{ij}^{B}$ if they buy. The unconditional
feasible set of utilities is therefore%
\[
\mathcal{F}_{ij}=\mathcal{F}_{ij}^{R}\cup \mathcal{F}_{ij}^{B}
\]%
which is depicted in figure~\ref{fig:ctu}. Here, the set $\mathcal{F}_{ij}^{R}\cup 
\mathcal{F}_{ij}^{B}$ is not convex, so its Pareto frontier, which is
represented in thick black on figure~\ref{fig:ctu}, has some points that are never
picked by the Pareto approach; these points are the points that do not 
coincide with the red envelope. Yet, which point is selected on the Pareto
frontier is determined endogenously at equilibrium by the other options that
are available to the partners in other potential match. This may perfectly
be a point which is not picked up by the Pareto weights method.

One may respond to that critique by assuming that agents have a von
Neumann-Morgenstern utility and lotteries over $\mathcal{F}_{ij}^{R}\cup 
\mathcal{F}_{ij}^{B}$, which means in economic terms that partners are
allowed to randomize between buying and renting. In that case, the set of
feasible utilities is replaced by its convex envelope $\mathcal{F}%
_{ij}^{conv}:=conv(\mathcal{F}_{ij}^{R}\cup \mathcal{F}_{ij}^{B})$, and all
points of the frontier of the convexified set can be picked up this time by
the Pareto weights approach. But that approach relies on a bold assumption:
that partners rely on randomization for big decisions (here, home ownership
decision, but the argument would transpose for other public good decisions 
within the household, such  as fertility, residential location, labor market participation, etc.), and that such a
randomization actually increases their utility. It is permitted to remain 
skeptical of the plausibility of this behavioural assumption.

\bigskip

Throughout this chapter, we shall therefore adopt the Pareto efficiency approach, but not the Pareto weights approach.
We will assume that the household's bargaining process leads the partners to a Pareto efficient solution, but we will not assume that they randomize their decisions, so we will not follow the Pareto weights approach. In our framework, the point of the Pareto frontier that will be picked is determined by competitive market equilibrium, and by the multiple outside options that each partner can consider getting with other potential partners. As we shall see, in the case of large markets, there is a unique such point.

\subsection{Properties of feasible sets}

\newcommand{\scalefig}{0.49}
\begin{figure}[t]
\caption{Examples of bargaining sets.}
\centering
\begin{subfigure}[b]{0.49\textwidth}
        
        \includegraphics[scale=\scalefig]{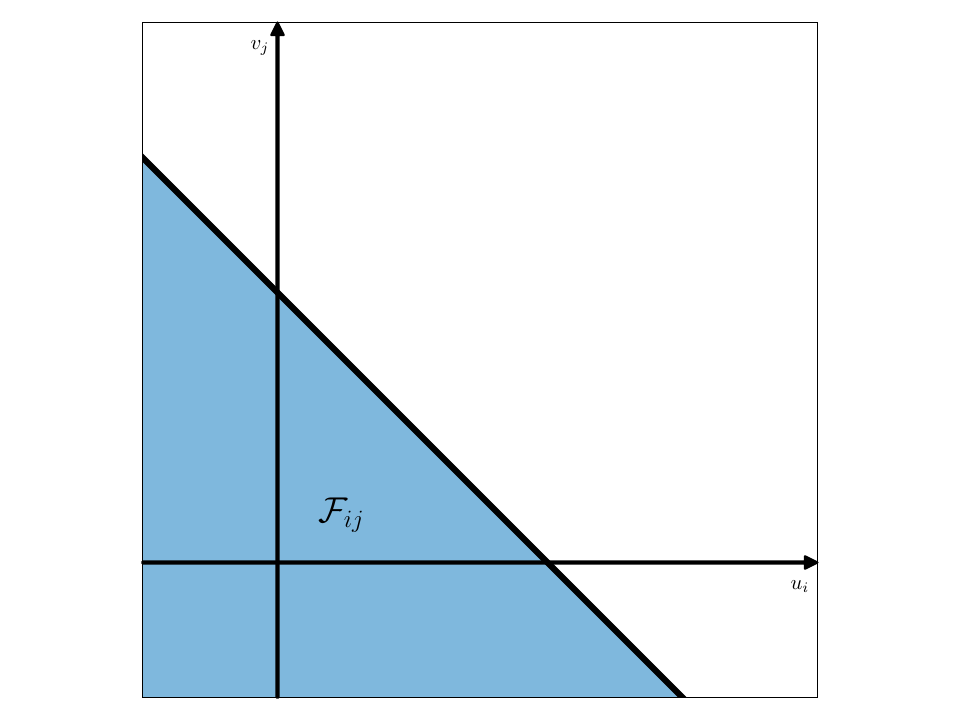}
                \caption{TU}\label{fig:TU}
    \end{subfigure}
\begin{subfigure}[b]{0.49\textwidth}
        \includegraphics[scale=\scalefig]{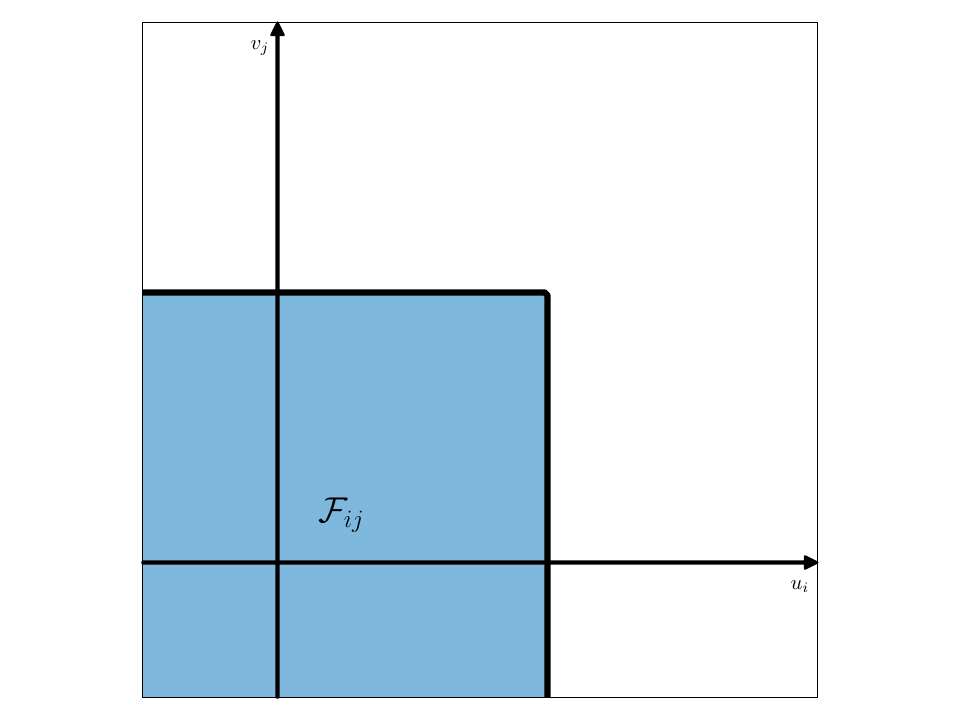}
        \caption{NTU}\label{fig:NTU}
    \end{subfigure}\newline
\begin{subfigure}[b]{0.49\textwidth}
        \includegraphics[scale=\scalefig]{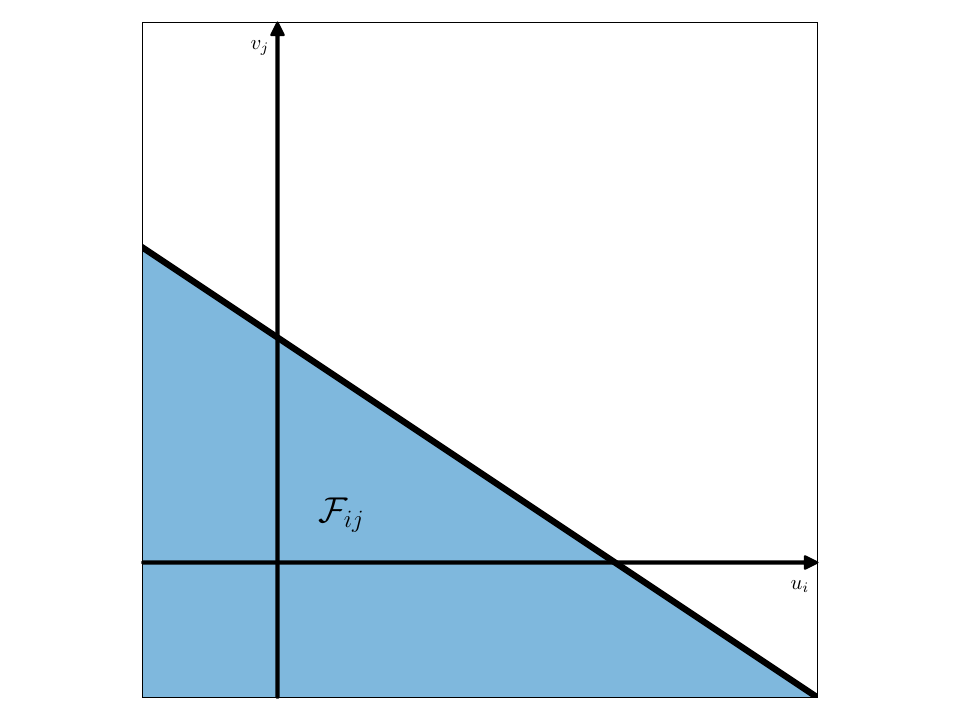}
                \caption{LTU}\label{fig:LTU}
    \end{subfigure}
\begin{subfigure}[b]{0.49\textwidth}
        \includegraphics[scale=\scalefig]{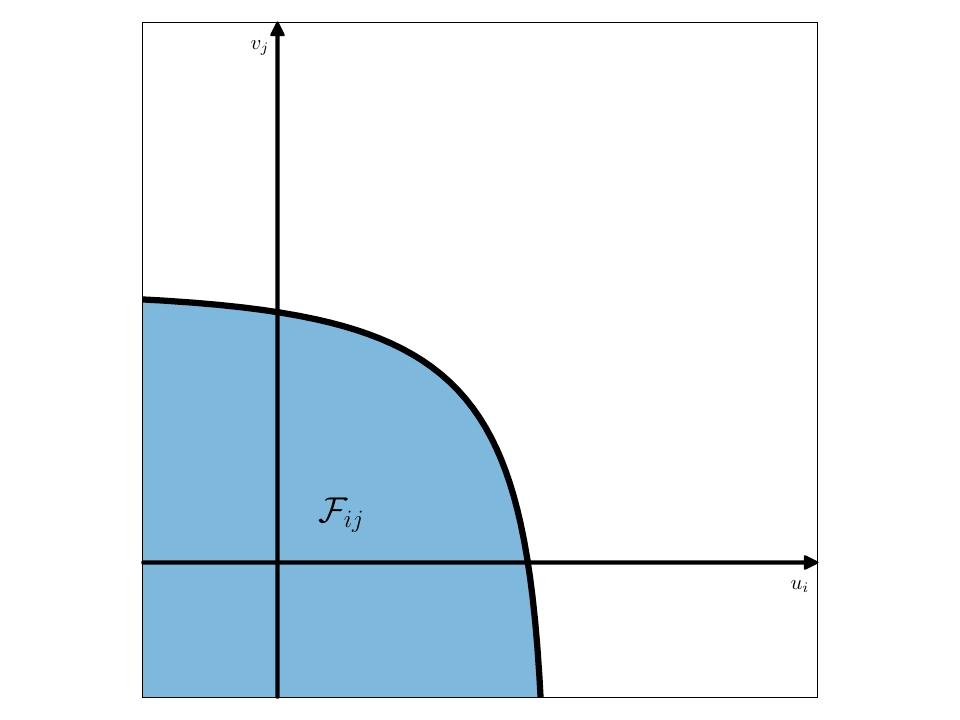}
        \caption{ETU}\label{fig:ETU}
    \end{subfigure}\newline
\begin{subfigure}[b]{0.49\textwidth}
        \includegraphics[scale=\scalefig]{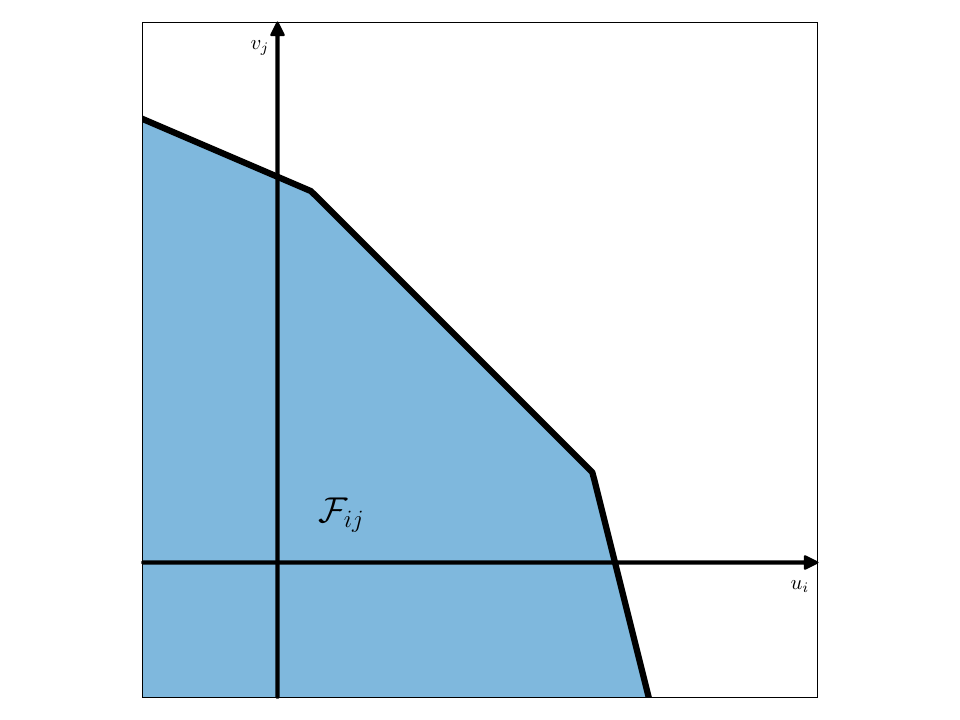}
        \caption{Intersection of LTU}\label{fig:LTUintersection}
    \end{subfigure}
\begin{subfigure}[b]{0.49\textwidth}
        \includegraphics[scale=\scalefig]{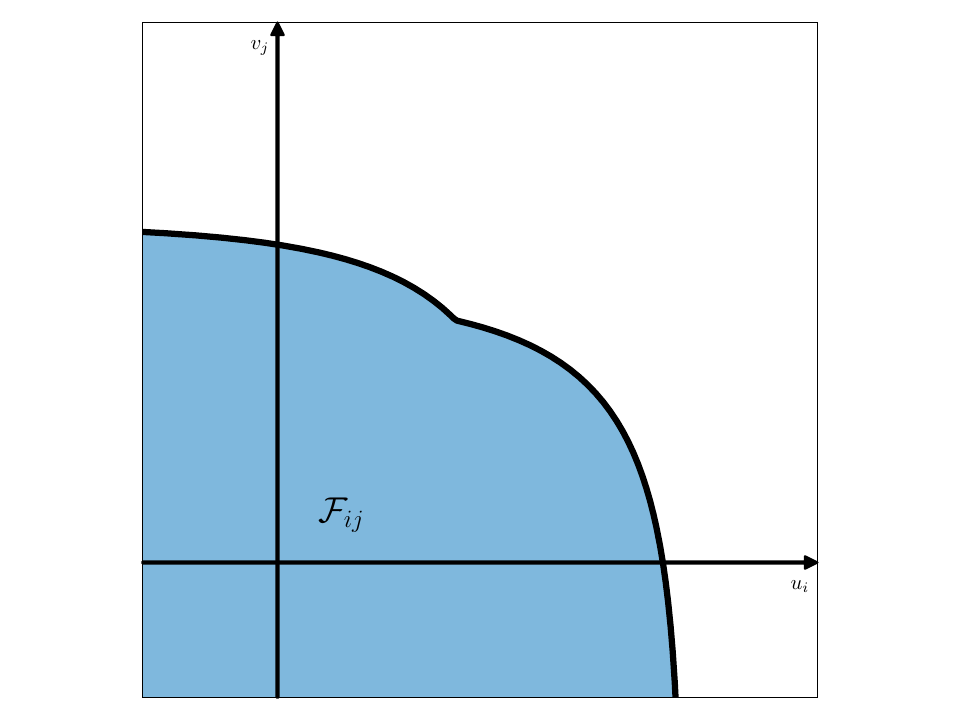}
        \caption{Union of ETU}\label{fig:ETUunion}
    \end{subfigure}

\label{fig:examples}
\end{figure}

In this section we define the precise requirements that we will impose on the feasible set of utilities. We suppose that if man $i\in \mathcal{I}$ and woman $j\in \mathcal{J}$ are matched, they bargain over utility allocations $( u_{i},v_{j}) \in
\mathcal{F}_{ij}$. We call the set $%
\mathcal{F}_{ij}$ a \textit{feasible utility set}, or \textit{bargaining set} and make the following assumptions on the structure of such sets:

\begin{definition}[\cite{galichonkominersweber2019}]
\label{def:bargainingsets}The set $\mathcal{F}_{ij}$ is a \emph{proper
bargaining set} if the following three conditions are met:

(i) $\mathcal{F}_{ij}$ is closed and nonempty.

(ii) $\mathcal{F}_{ij}$ is \emph{lower comprehensive}: if $u^{\prime }\leq u$%
, $v^{\prime }\leq v$, and $( u,v) \in \mathcal{F}_{ij}$, then $%
( u^{\prime },v^{\prime }) \in \mathcal{F}_{ij}$.

(iii) $\mathcal{F}_{ij}$ is \emph{bounded above}: If $u_{n}\rightarrow
+\infty $ and $v_{n}$ bounded below, then for $N$ large enough $(
u_{n},v_{n}) \notin \mathcal{F}$ for $n\geq N$; similarly for $u_{n}$
bounded below and $v_{n}\rightarrow +\infty $.
\end{definition}

Figure \ref{fig:examples} shows the bargaining sets associated to some of the examples explored in the introduction. As one can see on the one corresponding to non-transferable utility (NTU), any point on the frontier of the feasible set except for the corner is \emph{not} Pareto efficient because it is dominated by the corner. This will lead to interesting normative implications of the NTU model in terms of Pareto efficiency and equal treatment, to be discussed in section~\ref{par:normative-properties}. As in Nöldeke and Samuelson's chapter~\parencite{noeldeke_samuelson_handbook} in the present Handbook, the bargaining sets need not be convex in our framework. We only require the sets to be closed (which is necessary for efficient
allocations to exist) and non-empty, lower comprehensive (which is equivalent to free disposal), nonempty (which along with lower comprehensiveness implies that if both partners’ demands are low enough, they can always be fulfilled) and bounded above (which is a scarcity requirement, that rules out the possibility of both partners obtaining arbitrarily large utilities).

\FloatBarrier

\subsection{Distance-to-frontier function}

The distance-to-frontier function implicitly describes bargaining sets. It measures the signed distance of a utility allocation $(u,v)$ to the frontier of the bargaining set $\mathcal{F}_{ij}$ along the diagonal, up to a factor of $\sqrt{2}$. The distance is positive if the allocation lies outside the feasible set, negative if it lies inside the feasible set, and equal to 0 if it lies on the frontier. GKW formally define the distance-to-frontier function as follows:

\begin{definition}[Distance-to-frontier function]
\label{def:distance-to-frontier}
The distance-to-frontier function $D_{\mathcal{F}_{ij}}:\mathbb{R}%
^{2}\rightarrow \mathbb{R}$ of a proper bargaining set $\mathcal{F}_{ij}$ is
defined by%
\begin{equation}
D_{\mathcal{F}_{ij}}( u,v) =\min \{ z\in \mathbb{R}:(
u-z,v-z) \in \mathcal{F}_{ij}\} .  \label{defPsi}
\end{equation}
\end{definition}

Definition \ref{def:bargainingsets} ensures the distance-to-frontier function exists. In addition, as shown in GKW, it satisfies the following properties:

\begin{lemma}
\label{lm:propertiesOfPsi}Let $\mathcal{F}_{ij}$ be a proper bargaining set.
Then:
\renewcommand{\labelenumi}{(\roman{enumi})}
\begin{enumerate}
    \item $\mathcal{F}_{ij}=\{ ( u,v) \in \mathbb{R}%
^{2}:D_{\mathcal{F}_{ij}}( u,v) \leq 0\} $.
    \item For every $( u,v) \in \mathbb{R}^{2}$, $D_{\mathcal{F}%
_{ij}}( u,v) \in ( -\infty ,+\infty ) $.
    \item $D_{\mathcal{F}_{ij}}$ is $\gg $-isotone, meaning that $\ (
u,v) \leq ( u^{\prime },v^{\prime }) $ implies $D_{\mathcal{%
F}_{ij}}( u,v) \leq D_{\mathcal{F}_{ij}}( u^{\prime
},v^{\prime }) $; and $u<u^{\prime }$ and $v<v^{\prime }$ jointly
imply $$D_{\mathcal{F}_{ij}}( u,v) <D_{\mathcal{F}_{ij}}(
u^{\prime },v^{\prime }) .$$
    \item $D_{\mathcal{F}_{ij}}$ is continuous.
    \item $D_{\mathcal{F}_{ij}}( a+u,a+v) =a+D_{\mathcal{F}_{ij}}(
u,v) $.
\end{enumerate}
\end{lemma}

Representing bargaining sets implicitly with distance-to-frontier functions is particularly useful in practice. One important advantage of that approach is that basic geometric operations on bargaining sets can be expressed as algebraic operations on the corresponding distance-to-frontier functions. Specifically, the distance-to-frontier function for the union of sets is the minimum of the corresponding distance-to-frontier functions, and the distance-to-frontier function for the intersection of sets is the maximum of the distance-to-frontier functions. The following lemma, due to GKW, formally states this:

\begin{lemma}
\label{lem:geom}Assume that $\mathcal{F}_{1},\ldots,\mathcal{F}_{K}$ are $K$
proper bargaining sets, as introduced in definition~\ref{def:bargainingsets}.
Then:
\begin{enumerate}
    \item The sets $\mathcal{F}_{\cup }=\bigcup_{k\in \{ 1,\ldots,K\} }%
\mathcal{F}_{k}$ and $\mathcal{F}_{\cap }=\bigcap_{k\in \{
1,\ldots,K\} }\mathcal{F}_{k}$ are proper bargaining sets,
    \item The distance to frontier functions for the sets $\mathcal{F}_{\cup }=\bigcup_{k\in \{ 1,\ldots,K\} }%
\mathcal{F}_{k}$ and $\mathcal{F}_{\cap }=\bigcap_{k\in \{
1,\ldots,K\} }\mathcal{F}_{k}$ are given by%
\begin{align*}
D_{\mathcal{F}_{\cup }}(u,v) &=\min \{D_{\mathcal{F}_{1}}(u,v),\ldots,D_{%
\mathcal{F}_{K}}(u,v)\} \\
D_{\mathcal{F}_{\cap }}(u,v) &=\max \{D_{\mathcal{F}_{1}}(u,v),\ldots,D_{%
\mathcal{F}_{K}}(u,v)\}.
\end{align*}
\end{enumerate}
\end{lemma}

Using this lemma, we can construct complex models from elementary ones. For example, the bargaining set in example \ref{ex:introintersectionLTU} is the intersection between elementary bargaining sets of the same type as the ones described in example \ref{ex:introsimpleLTU}. Similarly, the bargaining set in example \ref{ex:introunionETU} is the union between elementary bargaining sets of the same type as the ones described in example \ref{ex:introsimpleETU}. 

\subsection{Examples}\label{sec:examples}

In this section, we revisit examples \ref{ex:introsimpleLTU}, \ref{ex:introintersectionLTU}, \ref{ex:introsimpleETU} and \ref{ex:introunionETU} presented in the introduction. But first, we discuss two cornerstone models in the matching literature, namely matching models with transferable utility and non-transferable utility, and explain how they fit into our framework.

\subsubsection{Transferable Utility models\label{ex:TU}}
In the classical TU matching model, it is assumed that the partners generate a \emph{joint surplus} $\Phi _{ij}=\alpha _{ij}+\gamma _{ij}$, which they split between themselves. Utility is perfectly transferable:\ if one partner gives up one
unit of utility, then the utility of the other partner increases exactly by one unit. The
Pareto efficient utility allocations $(u,v)$ are such that $u+v=\Phi _{ij}$.
The bargaining set is
\begin{equation}
\mathcal{F}_{ij}=\{ ( u,v) \in \mathbb{R}^{2}:u+v\leq \Phi
_{ij}\} ,  \label{TU-feasibleSet}
\end{equation}%
from which we derive the distance-to-frontier function
\begin{equation}
D_{ij}( u,v) =\frac{u+v-\Phi _{ij}}{2}.  \label{TU}
\end{equation}%
See figure \ref{fig:TU} for an illustration of a TU bargaining set.

\subsubsection{Non-Transferable Utility models\label{ex:NTU}}
In non-transferable utility models, utility is not
transferable at all. The maximum obtainable utility of each partner is
fixed and is independent from what the other partner gets. Assuming that when man $i$ and woman $j$ are matched they bring surpluses $\alpha _{ij}$ and $\gamma _{ij}$, respectively, the bargaining set is
\begin{equation}
\mathcal{F}_{ij}=\{ ( u,v) \in \mathbb{R}^{2}:u\leq \alpha
_{ij},v\leq \gamma _{ij}\} ,  \label{NTU-feasibleSet}
\end{equation}%
from which we derive the distance-to-frontier function
\begin{equation}
D_{ij}( u,v) =\max \{u-\alpha _{ij},v-\gamma _{ij}\}.  \label{NTU}
\end{equation}
See figure \ref{fig:NTU} for an illustration of a NTU bargaining set.

\subsubsection{Linearly Transferable Utility models}
Consider the setting of example \ref{ex:introsimpleLTU}, with one modification. We assume that besides having workers pay income taxes (recall that wages are taxed at a rate $\tau_{ij}$ on the workers' side), firms pay social contributions so wages are taxed at a rate $\kappa_{ij}$ on their side. We let the tax rates $\tau_{ij}$ and $\kappa_{ij}$ depend on both $i$ and $j$. In this context, if worker $i$ and firm $j$ match and set a wage $w_{ij}$, they receive utilities $u_{i}=\alpha _{ij}+(1-\tau_{ij})w_{ij}$ and $v_{j}=\gamma
_{ij}-(1+\kappa_{ij})w_{ij}$, respectively. The corresponding bargaining set is given by 
\begin{equation*}
\mathcal{F}_{ij}=\left\{ \left( u,v\right) \in \mathbb{R}^{2}:\lambda
_{ij}u+\zeta _{ij}v\leq \Phi _{ij}\right\} ,
\end{equation*}%
where $\lambda _{ij}=1/(1-\tau_{ij})>0$, $\zeta _{ij}=1/(1+\kappa_{ij})>0$, and $\Phi
_{ij}=\lambda _{ij}\alpha _{ij}+\zeta _{ij}\gamma _{ij}$. We obtain the distance-to-frontier function
\begin{equation}
D_{ij}\left( u,v\right) =\frac{\lambda _{ij}u+\zeta _{ij}v-\Phi _{ij}}{%
\lambda _{ij}+\zeta _{ij}}.  \label{LTU}
\end{equation}
Note that without loss of generality, we can normalize $\lambda _{ij}+\zeta _{ij} =2$, and when $\lambda _{ij} = \zeta _{ij} = 1$ we recover the standard TU model. See figure \ref{fig:LTU} for an illustration of a LTU bargaining set.

\subsubsection{Exponentially Transferable Utility models}
Consider the setting of example \ref{ex:introsimpleETU}. Recall that when man $i$ and woman $j$ match, they share income $B_{ij}$ which they allocate to private consumption for the man ($c_i$) and for the woman ($c_j$), so that $c_{i}+c_{j}=B_{ij}$. Let the man and the woman utility functions be given by
\begin{equation*}
u_{ij}( c_{i}) =\alpha _{ij} +\tau
_{ij}\log c_{i}\text{ and }v_{ij}( c_{j}) =\gamma
_{ij} +\tau _{ij}\log c_{j}
\end{equation*}%
where the terms $\alpha _{ij}$ and $\gamma _{ij}$ capture marital affinities (non-economic gains to marriage). In this case
the bargaining set is 
\begin{equation*}
\mathcal{F}_{ij}=\left\{ \left( u,v\right) \in \mathbb{R}^{2}:\exp \left( \frac{u-\alpha _{ij}}{\tau _{ij}}\right) +\exp
\left( \frac{v-\gamma _{ij}}{\tau _{ij}}\right) \leq B_{ij} \right\}.
\end{equation*}%

The expression of $D_{ij}$ is:%
\begin{equation}
D_{ij}( u,v) =\tau _{ij}\log \left( \frac{\exp \left( \frac{%
u-\alpha _{ij}}{\tau _{ij}}\right) +\exp \left( \frac{v-\gamma _{ij}}{\tau
_{ij}}\right) }{B_{ij}}\right).  \label{ETU}
\end{equation}%

As pointed out in GKW, an interesting property of this exponentially transferable utility model is that it interpolates between the
non-transferable and fully transferable utility models. Indeed, suppose that $B_{ij}=2$, then $\tau _{ij}\rightarrow 0$ gives the NTU model, while $\tau
_{ij}\rightarrow +\infty $ gives the TU model. Here, we can think of the parameter $%
\tau _{ij}$ as measuring the \textit{degree of
transferability}. See figure \ref{fig:ETU} for an illustration of a ETU bargaining set.

Note that after a nonlinear rescaling, the ETU model can be expressed as a LTU model. Let $\tau_{ij}=\tau$, for all pairs $(i,j)$, and let $U = \exp(u/\tau)$ and $V = \exp(v/\tau)$. The utility allocations $(U,V)$ on the Pareto frontier satisfy the equation $U \exp(-\alpha_{ij}/\tau) + V \exp(-\gamma_{ij}/\tau) = B_{ij}$. Letting $\lambda_{ij} = \exp(-\alpha_{ij}/\tau)$, $\zeta_{ij} = \exp(-\gamma_{ij}/\tau)$ and $\Phi_{ij} = B_{ij}$, the equation describing the frontier rewrites $\lambda
_{ij}U+\zeta _{ij}V= \Phi _{ij}$ with associated distance-to-frontier function 
\begin{equation*}
D_{ij}\left( U,V\right) =\frac{\lambda _{ij}U+\zeta _{ij}V-\Phi _{ij}}{%
\lambda _{ij}+\zeta _{ij}},
\end{equation*}
which is the LTU model.

\subsubsection{Revisiting example \ref{ex:introintersectionLTU}}

We now reexamine example \ref{ex:introintersectionLTU} in light of the LTU model and lemma \ref{lem:geom}. Recall that in that example, wages are taxed only on the workers' side, according to a convex tax schedule. 



Let us introduce the quantity $\alpha _{ij}^{k}$, which we define recursively as follows: we let $\alpha _{ij}^{0}=\alpha _{ij}$, and $\alpha _{ij}^{k+1}=\alpha
_{ij}^{k}+\left( 1-\tau ^{k}\right) \left( t^{k+1}-t^{k}\right) $. The utility of a worker receiving gross wage $w_{ij}$ is given by $%
u_{i}=\min \{ \alpha _{ij}^{k}+\left( 1-\tau ^{k}\right)
w_{ij},k=0,\ldots,K\} $, while the utility of the firm is simply $%
v_{j}=\gamma _{ij}-w_{ij}$. The bargaining set is%
\begin{equation*}
\mathcal{F}_{ij}=\left\{ \left( u,v\right) \in \mathbb{R}^{2}:\forall k\in
\left\{ 0,\ldots,K\right\} ,u\leq \alpha _{ij}^{k}+\left( 1-\tau
^{k}\right) \left( \gamma _{ij}-v\right) \right\}.
\end{equation*}%

Let $\mathcal{F}^k_{ij} = \{ (u,v) \in \mathbb{R}^{2}: u\leq \alpha _{ij}^{k}+( 1-\tau^{k}) (\gamma _{ij}-v_{j})\}$. We have $\mathcal{F}_{ij} = \cap_k \mathcal{F}^k_{ij}$, i.e. the bargaining set $\mathcal{F}_{ij}$ is the intersection of the elementary LTU sets $\mathcal{F}^k_{ij}$. Each of these sets has a frontier that is linear (but not necessarily of slope $-1$), so that the frontier of the set $\mathcal{F}_{ij}$ is piecewise linear. From lemma~\ref{lem:geom}, we obtain the distance-to-frontier function

\begin{equation}
D_{ij}\left( u,v\right) =\max_{k\in \left\{ 0,\ldots,K\right\} }\left\{ \frac{%
u-\alpha _{ij}^{k}+\left( 1-\tau ^{k}\right) \left( v-\gamma _{ij}\right) }{%
2-\tau ^{k}}\right\}.  \label{exprDconvexTaxSchedule}
\end{equation}%

See figure \ref{fig:LTUintersection} for an example in which the bargaining set is the intersection of elementary LTU bargaining sets.

\subsubsection{Revisiting example \ref{ex:introunionETU}}

We now make use of lemma \ref{lem:geom} to reexamine example \ref{ex:introunionETU}. We consider a similar setting to example \ref{ex:introunionETU}, but assume that partners must decide on some amount of public good consumption $g\in \mathcal{G}$, where $\mathcal{G}$ is a closed set that may be finite or not. In some applications, this set might be discrete, e.g. in models featuring fertility decisions. Given the chosen amount of public good consumption, we assume that the partners allocate the remaining income $B_{ij}(g)$ to the private consumption of the man $(c_i)$ and of the woman ($c_j$), so that $c_{i}+c_{j}=B_{ij}(g)$. Suppose that the utilities received by the man and the woman are given by%
\begin{equation*}
u_{ij}( c_{i},g) =\alpha _{ij}( g) +\tau
_{ij}\log c_{i}\quad\text{ and }\quad v_{ij}( c_{j},g) =\gamma
_{ij}( g) +\tau _{ij}\log c_{j},
\end{equation*}%
respectively. The bargaining set is
\begin{equation*}
\mathcal{F}_{ij}=\left\{ \left( u,v\right) \in \mathbb{R}^{2}:\exists g\in\mathcal{G}, \exp \left( \frac{u-\alpha _{ij}\left( g\right) }{\tau _{ij}}\right) +\exp
\left( \frac{v-\gamma _{ij}\left( g\right) }{\tau _{ij}}\right) \leq B_{ij}(g) \right\}.
\end{equation*}%

Let $\mathcal{F}_{ij}(g) = \{ (u,v) \in \mathbb{R}^{2}:\exp ((u-\alpha _{ij}(g))/\tau _{ij}) +\exp( (v-\gamma _{ij}(g))/\tau _{ij}) \leq B_{ij}(g)\}$. Note that $\mathcal{F}_{ij} = \cup_{g\in\mathcal{G}} \mathcal{F}_{ij}(g)$, that is, the bargaining set is the union of elementary ETU sets. From lemma~\ref%
{lem:geom}, we obtain the distance-to-frontier function
\begin{equation}
D_{ij}\left( u,v\right) =\min_{g\in \mathcal{G}}\left\{\tau _{ij}\log \left( \frac{%
\exp \left( \frac{u-\alpha _{ij}\left( g\right) }{\tau _{ij}}\right) +\exp
\left( \frac{v-\gamma _{ij}\left( g\right) }{\tau _{ij}}\right) }{%
B_{ij}\left( g\right) }\right) \right\}.  \label{DTFpublicGood}
\end{equation}

See figure \ref{fig:ETUunion} for an example in which the bargaining set is the union of elementary ETU bargaining sets.

\subsubsection{Matching with uncertainty\label{ex:uncertainty}}

Consider a household bargaining setting similar to that of example \ref%
{ex:introsimpleETU}, to which we add uncertainty, following the insight of~%
\textcite{chadeeeckhout2017}. When man $i$ and woman $j$ match, they share
income $\tilde{B}_{ij}\in \mathbb{R}$ which we assume to be stochastic. We
let $\tilde{c}_{i}\in \mathbb{R}$ and $\tilde{c}_{j}\in \mathbb{R}$ be the
contingent private consumptions of the man and the woman, respectively, and
we assume that the corresponding expected utilities are respectively $%
u=\alpha _{ij}+\mathbb{E}[U_{i}(\tilde{c}_{i})]$ and $v=\gamma _{ij}+\mathbb{%
E}[V_{j}(\tilde{c}_{j})]$, where the von-Neumann-Morgenstern utility functions $U_i$ and $V_j$ are concave. Thus, the bargaining set for that couple is 
\begin{equation*}
\mathcal{F}_{ij}=\left\{ 
\begin{array}{c}
(u,v)\in \mathbb{R}^{2}:u\leq \alpha _{ij}+\mathbb{E}[U_{i}(\tilde{c}_{i})]%
\text{ and }v\leq \gamma _{ij}+\mathbb{E}[V_{j}(\tilde{c}_{j})] \\ 
\text{for each }(\tilde{c}_{i},\tilde{c}_{j})\text{ such that }\tilde{c}_{i}+%
\tilde{c}_{j}=\tilde{B}_{ij}\text{ holds almost surely}%
\end{array}%
\right\} ,
\end{equation*}
and we see that as soon as $U_i$ or $V_j$ are strictly concave, the utility is no longer fully transferable as the upper envelope of $\mathcal F_{ij}$ is strictly concave.

\section{Stable matchings without heterogeneity}

\subsection{Setting}

In our matching framework, men and women can form heterosexual pairs or remain unmatched. If a man $i$ or a woman $j$ remain unmatched, they receive their reservation utility, denoted $\mathcal{U}_{i0}$ and $\mathcal{V}_{0j}$, respectively\footnote{In our framework, the agents' utilities do not depend directly on what happens elsewhere on the market, e.g. on the masses of matched pairs or the masses of singles. For models featuring such peer effects, see \textcite{mourifie2019} and \textcite{mourifiesiow2021} for example.}. If man $i$ and woman $j$ match, they bargain over a proper bargaining set $\mathcal{F}_{ij}$. The indirect payoffs of man $i$ and woman $j$ are denoted $u_{i}$ and $v_{j}$, respectively, and are determined at equilibrium. Finally, $\mu _{ij}$ denote the matching which, like the indirect payoffs, is determined at equilibrium.

\subsection{Stability}

In the imperfectly transferable utility framework, we recall GKW's definition of an individual equilibrium outcome as a triple $( \mu _{ij},u_{i},v_{j})$ that satisfies the following conditions:

\begin{definition}[Individual equilibrium outcome]
\label{def:ITU-none}The triple $( \mu _{ij},u_{i},v_{j}) _{i\in
\mathcal{I},j\in \mathcal{J}}$ is an \emph{individual equilibrium outcome}
if the following three conditions are met:
\begin{enumerate}
\renewcommand{\labelenumi}{(\roman{enumi})}
\item $\mu _{ij}\in \{ {0,1}\} $, $\sum_{j}\mu _{ij}\leq 1$ and $%
\sum_{i}\mu _{ij}\leq 1$;

\item for all $i$ and $j$, $D_{ij}( u_{i},v_{j}) \geq 0$, with
equality if $\mu _{ij}=1$;

\item $u_{i}\geq \mathcal{U}_{i0}$ and $v_{j}\geq \mathcal{V}_{0j}$, with
equality if $\sum_{j}\mu _{ij}=0$ and if $\sum_{i}\mu _{ij}=0$, respectively.
\end{enumerate}
The vector $( \mu _{ij}) _{i\in \mathcal{I},j\in \mathcal{J}}$ is
an \emph{individual equilibrium matching} if and only if there exists a pair
of vectors $( u_{i},v_{j}) _{i\in \mathcal{I},j\in \mathcal{J}}$
such that $( \mu ,u,v) $ is an individual equilibrium outcome.
\end{definition}

Condition (i) is simply a feasibility condition on the matching $\mu$. Together, the inequalities in conditions (ii) and (iii) form the pairwise stability conditions. Specifically, at equilibrium, $%
D_{ij}( u_{i},v_{j}) \geq 0$ should hold for every $i$ and $j$. Suppose not: then there would be a pair $( i,j) $
such that $( u_{i},v_{j}) $ lies strictly within the feasible set $\mathcal{F}_{ij}$. Thus, there would exist payoffs $( u^{\prime },v^{\prime }) $ such that
$u^{\prime }\geq u_{i}$ and $v^{\prime }\geq v_{j}$ (with at least one strict
inequality) and $( u^{\prime },v^{\prime }) \in \mathcal{F}_{ij}$. This would imply that $i$ and $j$ could both be better off by matching together. Additionally, $u_{i}\geq \mathcal{U}_{i0}$ and $v_{j}\geq
\mathcal{V}_{0j}$ should hold for all $i$ and $j$, since partners can always leave an arrangement if it yields less than their reservation utility. Finally, note that if $i$ and $j$ are matched, we require $(
u_{i},v_{j}) $ to be feasible, that is $D_{ij}( u_{i},v_{j})
\leq 0$, hence this equation should hold with equality.  Similarly, if $i$ ($j$) is unmatched, then $u_{i}= \mathcal{U}_{i0}$ ($v_{j}=
\mathcal{V}_{0j}$). These equalities are known as the complementary slackness conditions.

The stability conditions in definition \ref{def:ITU-none} form the nonlinear counterpart to the usual stability conditions from the TU framework. Indeed, in that case recall that $D_{ij}( u,v) =\frac{u+v-\Phi _{ij}}{2}$, hence the inequalities from condition (ii) read as $u_{i}+v_{j}\geq
\Phi _{ij}$ for all $i$ and $j$, the well-known stability conditions met in the chapter of \textcite{salanie_handbook} in this Handbook.

\section{Stable matchings with heterogeneity}
\label{sec:stabilitywithhet}

We now extend our model by assuming that agents can be grouped into a finite number of observable types, and have heterogeneous (unobserved) tastes. This macroscopic analog of the individual equilibrium, called the ``aggregate equilibrium'', describes the equilibrium matching patterns and systematic payoffs for each type. We will define this concept in more detail below.

\subsection{Unobserved heterogeneity\label{par:setting}}

Within a given type, agents have similar observable characteristics, but heterogeneous tastes.
Let $x_i\in \mathcal{X}$ denote the observable type of man $i$ and $y_j\in \mathcal{Y}$ denote the observable type of woman $j$ (see notations in the introduction to this chapter). We assume that the feasible set of utilities $\mathcal{F%
}_{ij}$ jointly obtainable by $i$ and $j$ is a random set whose stochasticity
has the following structure.

\begin{assumption}
\label{ass:prefsheterog}There exist families of probability\
distributions $(\mathbf{P}_{x})_{x\in \mathcal{X}}$ and $(\mathbf{Q}%
_{y})_{y\in \mathcal{Y}}$ such that 
\begin{enumerate}
\renewcommand{\labelenumi}{(\roman{enumi})}
    \item if $i$ and $j$ are matched, there exists $(
U_{i},V_{j}) \in \mathcal{F}_{x_{i}y_{j}}$, where $\mathcal{F}_{x_{i}y_{j}}$ is a proper bargaining set in the sense of definition \ref{def:bargainingsets}, such that $%
u_{i}=U_{i}+\varepsilon _{iy_{j}}$ and $v_{j}=V_{j}+\eta _{x_{i}j}$,

    \item if $i$ and $j$ remain
single, then they obtain utilities $\varepsilon _{i0}$ and $\eta _{0j}$, respectively,
\end{enumerate}
where the random vectors $( \varepsilon _{iy}) _{y\in \mathcal{Y}%
_{0}}$ and $( \eta _{xj}) _{x\in \mathcal{X}_{0}}$ are i.i.d.\
draws from $\mathbf{P}_{x}$ and $\mathbf{Q}_{y}$, respectively.
\end{assumption}

Note that in the TU case, which is discussed in detail in the chapter of \textcite{salanie_handbook} in the present volume, assumption \ref{ass:prefsheterog} boils down to assuming that the joint surplus $\Phi
_{ij}$ can be decomposed into $\Phi _{ij}=\Phi
_{x_{i}y_{j}}+\varepsilon _{iy_{j}}+\eta _{x_{i}j}$, which is the
\textquotedblleft additive separability\textquotedblright\ assumption commonly found in the literature.

Next, we impose the following assumption on the
distributions of the idiosyncratic terms $(\varepsilon _{iy})_{y\in \mathcal{%
Y}_{0}}$ and $(\eta _{xj})_{x\in \mathcal{X}_{0}}$, $\mathbf{P}_{x}$ and $\mathbf{Q}_{y}$:

\begin{assumption}
\label{ass:Distrib}$\mathbf{P}_{x}$ and $\mathbf{Q}_{y}$ have non-vanishing
densities on $\mathbb{R}^{\mathcal{Y}_{0}}$ and $\mathbb{R}^{\mathcal{X}%
_{0}} $.
\end{assumption}

Assumption~\ref{ass:Distrib} requires the distributions $\mathbf{P}_{x}$ and $\mathbf{Q}_{y}$ to have full support and be absolutely continuous. Full-support ensures that for any pair of types $x$ and $y$, there exist individuals of those types with arbitrarily large valuations for each other. This implies that at equilibrium, there will be matches between individuals of all observable pairs of types. Absolute continuity, on the other hand, ensures that with probability one, both the men and women choice problems have unique solutions.

\subsection{Another parameterization of bargaining sets}

Before delving into the concept of aggregate equilibrium and its properties, we introduce an explicit parametrization of the bargaining frontier, which will prove instrumental in our analysis.

Let $\mathcal{F}_{xy}$ be a proper bargaining set and $D_{xy}$ be the associated distance-to-frontier function. Let $(u,v)$ be a utility allocation lying on the frontier of the bargaining set, i.e. such that $D_{\mathcal{F}%
_{xy}}(u,v) =0$. GKW define the \emph{wedge} $w$ as the
difference $w=u-v$. They show that there exists two 1-Lipschitz functions $\mathcal{U}_{\mathcal{F}_{xy}}$ and $\mathcal{%
V}_{\mathcal{F}_{xy}}$ defined on a nonempty open interval $(%
\underaccent{\bar}{w}_{xy},\bar{w}_{xy})$ such that $\mathcal{U}_{\mathcal{F}%
_{xy}}$ is nondecreasing and $\mathcal{V}_{\mathcal{F}_{xy}}$ is
nonincreasing, and such that the set of utility allocations that are located on the frontier of the bargaining set is given by $\{(\mathcal{U}_{\mathcal{%
F}_{xy}}( w) ,\mathcal{V}_{\mathcal{F}_{xy}}( w)
):w\in (\underaccent{\bar}{w}_{xy},\bar{w}_{xy})\}$. Therefore, these functions provide an explicit representation of the bargaining frontier.

In addition, GKW show that $\mathcal{U%
}_{\mathcal{F}_{xy}}( w) $ and $\mathcal{V}_{\mathcal{F}%
_{xy}}( w) $ are the unique values of $u$ and $v$ solving the equations $D_{\mathcal{F}_{xy}}( u,v) =0\text{ and }w=u-v$ and that they are given by
\begin{equation}
\mathcal{U}_{\mathcal{F}_{xy}}( w) =-D_{_{\mathcal{F}%
_{xy}}}( 0,-w), \text{ and }\mathcal{V}_{\mathcal{F}_{xy}}(
w) =-D_{_{\mathcal{F}_{xy}}}( w,0) .
\label{UcalAndVcalFromD}
\end{equation}

Building on this explicit wedge parameterization, we introduce the following technical restriction on the bargaining sets $\mathcal{F}_{xy}$:

\begin{assumption}
\label{ass:bargainingsets}The sets $\mathcal{F}_{xy}$ are such that for each
man type $x\in \mathcal{X}$, either all the $\bar{w}_{xy}$, $y\in \mathcal{Y}
$ are finite, or all the $\bar{w}_{xy}$, $y\in \mathcal{Y}$ coincide with $%
+\infty $. For each woman type $y\in \mathcal{Y}
$, either all the $\underaccent{\bar}{w}_{xy}$, $x\in \mathcal{X}$ are
finite, or all the $\underaccent{\bar}{w}_{xy}$, $x\in \mathcal{X}$ coincide
with $-\infty $.
\end{assumption}

This technical assumption, which is met in all the examples we considered, is helpful to prove the existence of an equilibrium; it imposes that the maximum
utility any man or woman can obtain with any partner is either always finite, or always infinite.

\subsection{Aggregate equilibrium}
\label{sec:aggreq}

We define an aggregate equilibrium, following GKW, as:

\begin{definition}[Aggregate equilibrium outcome]
\label{def:equil}The triple $( \mu _{xy},U_{xy},V_{xy}) _{x\in
\mathcal{X},y\in \mathcal{Y}}$ is an \emph{aggregate equilibrium outcome} if
the following three conditions are met:

(i) $\mu $ is an interior matching, i.e. $\mu \in \mathcal{M}^{\interior}$;

(ii) $(U,V)$ is feasible, i.e.%
\begin{equation}
D_{xy}( U_{xy},V_{xy}) =0,~\forall x\in \mathcal{X},y\in \mathcal{%
Y};  \label{feasibility2}
\end{equation}

(iii) $\mu $, $U$, and $V$ are related by the market-clearing condition
\begin{equation}
\mu =\nabla G( U) =\nabla H( V)  \label{mktClearing}
\end{equation}
where
\begin{equation}
G( U) =\sum_{x\in \mathcal{X}}n_{x}\mathbb{E}\left[ \max_{y\in
\mathcal{Y}}\left\{ U_{xy}+\varepsilon _{iy},\varepsilon _{i0}\right\} %
\right] \quad \text{and}\quad H(V)=\sum_{y\in \mathcal{Y}}m_{y}\mathbb{E}%
\left[ \max_{x\in \mathcal{X}}\left\{ V_{xy}+\eta _{xj},\eta _{0j}\right\} %
\right]  \label{GandH}
\end{equation}
are the total indirect surplus of men and women, respectively. The vector $( \mu _{xy}) _{x\in \mathcal{X},y\in \mathcal{Y}}$ is
an \emph{aggregate equilibrium matching} if and only if there exists a pair
of vectors $( U_{xy},V_{xy}) _{x\in \mathcal{X},y\in \mathcal{Y}}$
such that $( \mu ,U,V) $ is an aggregate equilibrium outcome.
\end{definition}

Implicitly, the justification for using definition \ref{def:equil} lies in the fact that we are looking for an individual equilibrium $( \mu _{ij},u_{i},v_{j})$
with the property that there exist two vectors of \emph{systematic utilities} $( U_{xy}) $ and $%
( V_{xy}) $ such that if $i$ is matched with $j$, then $%
u_{i}=U_{x_{i}y_{j}}+\varepsilon _{iy_{j}}$, and $v_{j}=V_{x_{i}y_{j}}+\eta
_{x_{i}j}$. Although it might seem restrictive at first glance, GKW demonstrate that: (i) an individual equilibrium of this form always exists, and
(ii) under the additional assumption that the frontier of the feasible set is strictly downward sloping (which rules out the NTU case, for example),  \emph{any} individual equilibrium is of this form. We expand on these ideas in remark \ref{rk:eqtransfers} below. 

\bigskip 

Let us now examine each of the conditions that make up definition \ref{def:equil}. Condition (i) states that the matching must be feasible and lie in the strict interior of $\mathcal{M}$, i.e. that $\mu _{xy}>0$, $ \sum_{y\in \mathcal{Y}}\mu _{xy}<n_{x}$ and $\sum_{x\in \mathcal{X}}\mu
_{xy}<m_{y}$.  Condition (ii) relates the systematic utilities  $( U_{xy}) $ and $%
( V_{xy}) $ through a feasibility condition. To understand where these systematic utilities come from, let us adapt the definition of an individual equilibrium to our framework with parametrized heterogeneity. The triple $( \mu
_{ij},u_{i},v_{j}) $ is an individual equilibrium outcome when: $\mu _{ij}\in \{ {0,1}\} $, $\sum_{j}\mu _{ij}\leq 1$ and $%
\sum_{i}\mu _{ij}\leq 1$; for all $i$ and $j$, $D_{x_{i}y_{j}}( u_{i}-\varepsilon_{iy_{j}},v_{j}-\eta _{x_{i}j}) \geq 0$, with equality if $\mu _{ij}=1$; and $u_{i}\geq \varepsilon _{i0}$ and $v_{j}\geq \eta _{0j}$ with equality if $\sum_{j}\mu _{ij}=0$ and $\sum_{i}\mu _{ij}=0$, respectively. Note that the nonlinear stability conditions read: for any pair of types $x\in \mathcal{X}$
and $y\in \mathcal{Y}$,%
\begin{equation*}
\min_{\substack{ i:x_{i}=x  \\ j:y_{j}=y}}\{D_{xy}( u_{i}-\varepsilon
_{iy},v_{j}-\eta _{xj}) \}\geq 0,
\end{equation*}%
which holds with equality if there is a matching between a man of type $x$ and a woman
of type $y$. Let us define $U_{xy}=\min_{i:x_{i}=x}\{
u_{i}-\varepsilon _{iy}\} $ and $V_{xy}=\min_{j:y_{j}=y}\{
v_{j}-\eta _{xj}\} $; then the above inequality yields $D_{xy}( U_{xy},V_{xy}) \geq 0$. GKW show that, under weak conditions, this actually holds as an equality, that is
\begin{equation}
D_{xy}( U_{xy},V_{xy}) =0.  \label{intFeas}
\end{equation}

Condition (iii) relates the matching and the systematic utilities through market clearing conditions. It builds on the intuition that the matching problem with heterogeneity in tastes is analogous to a pair of discrete choice problems, each pertaining to one side of the market. The discrete choice formulation allows us to relate the vector of systematic utilities $( U_{xy}) $ and $%
( V_{xy}) $ to the equilibrium matching $\mu $ such that $\mu
_{xy}$ is the mass of men of type $x$ and women of type $y$ who mutually prefer each other. Note that from the definition of $U_{xy}$ and $V_{xy}$, we have $%
u_{i}\geq \max_{y\in \mathcal{Y}}\{ U_{xy}+\varepsilon
_{iy},\varepsilon _{i0}\} $ and $v_{j}\geq \max_{x\in \mathcal{X}%
}\{ V_{xy}+\eta _{xj},\eta _{0j}\} $, which can be shown to hold with equality. Thus we obtain the discrete choice problems
\begin{equation*}
u_{i}=\max_{y\in \mathcal{Y}}\{ U_{x_{i}y}+\varepsilon
_{iy},\varepsilon _{i0}\} \text{ and }v_{j}=\max_{x\in \mathcal{X}%
}\{ V_{xy_{j}}+\eta _{xj},\eta _{0j}\}
\end{equation*}%
in which each agent chooses the type of their partner. To obtain the mass of men of type $x$ demanding a partner of type $y$ and women of type $y$ demanding men of type $x$, we refer to the Daly-Zachary-Williams theorem \parencite{dalyzachary1978, williams1977} which indicates that these masses are given by $\partial G( U)
/\partial U_{xy}$ and $ \partial H( U)
/\partial V_{xy}$, respectively. At equilibrium, these quantities should coincide for all couple types, that is, $\nabla
G(U)=\nabla H( V) $ in vector notation. 

\begin{remark}[Equilibrium transfers and observable types]\label{rk:eqtransfers} The fact that at equilibrium the transfers only depend on observable types is a property of the model, not an assumption, and follows from the ``separability assumption,'' which is well-known in the literature on transferable utility following \textcite{choosiow2006}, and was highlighted in subsequent works by \textcite{chiapporisalanieweiss2017} and \textcite{galichonsalanie2022}. The key message of section \ref{sec:stabilitywithhet} is that there is no difference between TU and ITU when it comes to the implications of this assumption (suitably adapted). In both cases, as soon as we can show that there is an equilibrium that only depends on observable types, then that is the only equilibrium. The intuition is that it does not matter who an individual pays the equilibrium price to ; any attempt to use the information in $\epsilon$ (or $\eta$) to change that price would go against that individual's interests or their partner's, and therefore against the notion of equilibrium.
    
\end{remark}

\subsection{Existence, uniqueness and computation}\label{sec:existenceuniqueness}

GKW propose to reformulate the matching model as a demand system to show existence and uniqueness of an aggregate equilibrium outcome $( \mu ,U,V) $. Recall that the condition $D_{xy}( U_{xy},V_{xy})
=0 $ in definition \ref{def:equil} is equivalent to the existence of wedges $W_{xy} = U_{xy}-V_{xy}$ such that $U_{xy} = \mathcal{U}_{xy}( W_{xy}) $ and $V_{xy} = \mathcal{V}_{xy}(
W_{xy}) $. From now on, these wedges will play the role of market prices while the couple types $xy\in\mathcal{XY}$ will be treated as goods that are being produced by men and demanded by women. Interpreting $\partial G( \mathcal{U}( W) ) /\partial
U_{xy}$ and $\partial H(
\mathcal{V}( W) ) /\partial V_{xy}$ as the supply and demand of the $xy$ good, respectively, we introduce the excess demand function:

\begin{equation}
Z( W) :=\nabla H( \mathcal{V}( W) ) -\nabla
G( \mathcal{U}( W) ).  \label{excessDemand}
\end{equation}%
The next step is to find a price vector $W$ such that the excess demand is zero for all goods. GKW show that such a price vector exists and is unique:

\begin{theorem}[\cite{galichonkominersweber2019}]
\label{thm:existenceUniqueness}Under assumptions \ref{ass:prefsheterog}, \ref{ass:Distrib} and \ref%
{ass:bargainingsets}, there exists a unique vector $W$
such that%
\begin{equation}
Z( W) =0.  \label{zeroED}
\end{equation}
\end{theorem}

Finally, we can obtain the unique aggregate equilibrium outcome $( \mu ,U,V) $, where $\mu $, $%
U $, and $V$ are related to the solution $W$ to system (\ref{zeroED}) by $%
U_{xy}=\mathcal{U}_{xy}( W_{xy}) $, $V_{xy}=\mathcal{V}%
_{xy}( W_{xy}) $, and $\mu =\nabla G( U) =\nabla
H( V) $ (Corollary 1, GKW).

\bigskip
As the proofs of existence and uniqueness in theorem \ref{thm:existenceUniqueness} are instructive, they are succinctly discussed below.

The proof of existence is constructive, and relies on a Jacobi algorithm. The algorithm's ability to produce a solution crucially hinges on the excess demand function $Z$ satisfying the \emph{%
gross substitutability} condition (see GKW, Proposition 3). We now explain intuitively what that condition means in our setting, by describing what happens when the price $W_{xy}$ of good $xy$ increases (keeping all other prices constant). First, excess demand for good $xy$ decreases as there is less demand from women and more supply from men for that good. Second, for women of type $y$, other types of men become relatively more attractive, hence excess demand for goods $x^\prime y$ (where $x^\prime\neq x$) increases. Conversely, for men of type $x$, other types of women become relatively less attractive, hence excess demand for goods $xy^\prime$ (where $y^\prime\neq y$) increases. Third, the excess demand for good $x^\prime y^\prime$ does not respond to changes in the price of good $xy$ if $x^\prime\neq x$ and $y^\prime\neq y$. Fourth, as $W_{xy}$ increases, singlehood becomes weakly less attractive for all men, and strongly less so for men of type $x$, while singlehood becomes more attractive for women. Hence, the excess demand, aggregated over all couple types, decreases. 

The algorithm then works as follows. We start by choosing a vector $W^{0}$ of initial prices, that are high enough so that excess demand is negative for all couple types $xy$ (i.e. so that $Z( W^{0}) \leq 0$). Then, at each step $t\geq1$, we decrease the price of each good $xy$ such that the excess demand for that good is zero, keeping all other prices fixed (i.e. we set $W_{xy}^{t}$ to be such that $Z(W_{xy}^{t},W_{-xy}^{t-1}) =0$). Gross substitutability ensures that, for all pairs $xy$, the $W_{xy}^{t}$ are decreasing while the excess demand remains negative at every step. GKW show that the sequence $W_{xy}^{t}$ remains
bounded below; thus, it converges monotonically and its limit is a solution to equation \eqref{zeroED}. The algorithm is described formally below:

\begin{algorithm}
\ \newline
\label{algo:jacobi}
\begin{tabular}{r|p{5in}}
Step $0$ & Start with $W^{0}$ such that $Z( W^{0}) \leq 0$. \\
Step $t$ & For each $x\in \mathcal{X},y\in \mathcal{Y}$, let $W_{xy}^{t}$ solve $Z(W_{xy}^{t},W_{-xy}^{t-1}) =0$.
\end{tabular}

\smallskip
\noindent The algorithm terminates when $\sup_{xy\in \mathcal{X}\times \mathcal{Y}%
}|W_{xy}^{t}-W_{xy}^{t-1}|<\epsilon $.
\end{algorithm}

Uniqueness follows directly from the results of \textcite{berrygandhihaile2013}, which imply that $Z$ is inverse isotone. By definition, the mapping $Z$ is inverse isotone if, for any $W$ and $W^\prime$, $Z( W) \leq Z(W^\prime)$ implies $W \leq W^\prime$. The following argument established uniqueness. Suppose there are two vectors $W$ and $W^\prime$ such that $Z( W) =Z(W^\prime) =0$. Since  $Z( W) \leq Z(W^\prime)$, then $W\leq W^\prime$. And since $Z( W) \geq Z(W^\prime)$, then $W\geq W^\prime$. Hence $W=W^\prime$. 

\section{The ITU-logit model}

In this section and the next, we shall work under the assumption of logit heterogeneities. That is, we assume that $\mathbf{P}_{x}$ and $\mathbf{Q}_{y}$ are the distributions of i.i.d.~Gumbel (standard type~I extreme value) random variables.

While the logit assumption is not required for empirical work, it leads to a very tractable empirical framework. We will first show how our model can be conveniently reformulated as a matching function equilibrium model with partial assignment, and then examine some examples. In the next section, we will show how to estimate these models.

\begin{remark}[Beyond the logit case] While the assumption of logit heterogeneities features prominently throughout the remainder of this chapter, it is possible to use the framework with more general heterogeneities, albeit at the cost of convenience and tractability. 

We now provide a concise overview of the steps involved in determining the aggregate equilibrium under general heterogeneities. For a given price vector $W$, we compute the systematic utilities using the functions $\mathcal{U}(W)$ and $\mathcal{V}(W)$. We then draw many values of the idiosyncratic terms $(\varepsilon _{iy})_{y\in \mathcal{%
Y}_{0}}$ and $(\eta _{xj})_{x\in \mathcal{X}_{0}}$ from the desired distributions (provided that assumption \ref{ass:Distrib} is satisfied) and compute $H( \mathcal{V}( W) )$ and $G( \mathcal{U}( W) )$. Recall that in our demand system interpretation, $\partial G( \mathcal{U}( W) ) /\partial
U_{xy}$ is the supply of the $xy$ good, and $\partial H(
\mathcal{V}( W) ) /\partial V_{xy}$ is the
demand for that good at price vector $W$. We repeat these steps for possibly many values of $W$ until we find the price vector such that excess demand is zero for all goods on the market (see theorem \ref{thm:existenceUniqueness} and surrounding discussions). 
    
\end{remark}

\subsection{Matching functions\label{par:exITUlogit}}

With logit heterogeneities, it is well-known that the systematic utilities $U_{xy}$ and $V_{xy}$ can be recovered, in closed form, from the matching $\mu$. We have, by the classical log-odds-ratio formula, that $ U_{xy}=\log (\mu _{xy}/\mu _{x0})$ and $ V_{xy}=\log (\mu _{xy}/\mu _{0y})$, which, together with the feasibility condition $D_{xy}( U_{xy},V_{xy}) =0$ yields a relation between $\mu_{xy}$, $\mu_{x0}$ and $\mu_{0y}$, namely
\begin{equation}
D_{xy}( \log \mu _{xy}-\log \mu _{x0},\log \mu _{xy}-\log \mu
_{0y}) =0
\end{equation}
which implies $\log \mu _{xy}=-D_{xy}( -\log \mu _{x0},-\log \mu _{0y})$ by the translation property of distance-to-frontier functions. This gives us 
\begin{equation}
\mu _{xy}=M_{xy}(\mu_{x0},\mu_{0y}),  \label{eq:matching_function_def}
\end{equation}
where $M_{xy}$ is the \emph{matching function} associated to the ITU-logit model:
\begin{equation}
M_{xy}( \mu _{x0},\mu _{0y}) =\exp ( -D_{xy}( -\log \mu
_{x0},-\log \mu _{0y}) ) .  \label{matchingFunction}
\end{equation}
Although we call that relation a ``matching function'', it is important to note that this relationship between the number of ``vacancies'' $\mu_{x0}$ and $\mu_{0y}$ and the number of matches $\mu_{xy}$ does not exactly play the same role as matching functions that appear in the macroeconomic approaches to the labor markets, specifically in the search-and-matching literature, as surveyed in \textcite{petrongolopissarides2001}. There, a matching function specifies a dynamic relationship between matches and vacancies, with the understanding that the numbers of matches that are formed at the end of a period are a function of the number of vacancies at the start of the period. In our setting, we are at equilibrium, and the relation between these quantities is to be understood as a simultaneous equation between these quantities. Of course, it is possible to bring these approaches closer to each other when considering stationary equilibria in a search-and-matching market, as sketched in section~\ref{sec:search-and-matching}.

\subsection{Examples}

We revisit some of the examples introduced earlier in the special case of logit heterogeneity. Recall that the matching function is given by expression (\ref{matchingFunction}). By construction, the aggregate matching functions in this framework are homogeneous of degree 1 in the masses of singles.

In the TU model, the matching function becomes
\begin{equation}
M_{xy}( \mu _{x0},\mu _{0y}) =\mu _{x0}^{1/2}\mu _{0y}^{1/2}\exp
\frac{\Phi _{xy}}{2}.  \label{TU-logit}
\end{equation}%

The NTU specification with logit heterogeneities yields the matching function
\begin{equation}
M_{xy}( \mu _{x0},\mu _{0y}) =\min \{ \mu _{x0}e^{\alpha
_{xy}},\mu _{0y}e^{\gamma _{xy}} \}.  \label{NTU-logit}
\end{equation}

In the elementary LTU model explored in section \ref{sec:examples}, the matching function is
\begin{equation}
M_{xy}\left( \mu _{x0},\mu _{0y}\right) =
\mu _{x0}^{\frac{\lambda _{xy}}{\lambda _{xy} + \zeta _{xy}}} 
\mu _{0y}^{\frac{\zeta _{xy}}{\lambda _{xy} + \zeta _{xy}}} \exp\left(\frac{\Phi_{xy}}{\lambda _{xy} + \zeta _{xy}}\right).  \label{LTU-logit}
\end{equation}%
Note that we recover the TU matching function whenever $\lambda _{xy} = \zeta _{xy} = 1$.

Finally, in the case of the elementary ETU model with logit heterogeneity, we obtain the matching
function:%
\begin{equation}
M_{xy}\left( \mu _{x0},\mu _{0y}\right) =\left( \frac{\exp(-\alpha _{xy}/\tau
_{xy})\mu _{x0}^{-1/\tau _{xy}}+\exp(-\gamma _{xy}/\tau _{xy})\mu
_{0y}^{-1/\tau _{xy}}}{B_{xy}}\right) ^{-\tau _{xy}}.  \label{ETU-logit}
\end{equation}

As noted earlier, when $B_{xy}=2$ and $\tau _{xy}\rightarrow 0$, we recover the NTU-logit matching function, while $\tau _{xy}\rightarrow +\infty $, %
equation \eqref{ETU-logit} gives the TU-logit matching function.

\subsection{Matching function equilibrium}
\label{par:eqITU}

Matching functions relate $\mu _{xy}$ to the masses of singles $\mu _{x0}$ and $\mu _{0y}$. Combined with the requirement that $\mu \in \mathcal{M}^{\interior}$, this yields a system of nonlinear equations that fully characterize the aggregate matching equilibrium. The concept of matching function equilibrium builds on this idea. Following~\textcite{chenchoogalichonetal2023b}, we define:

\begin{definition}[Matching function equilibrium]\label{def: AMF} The matching patterns $\mu_{xy},\mu_{x0},\mu_{0y}$ form a \emph{matching function equilibrium}, if these quantities are related together by~\eqref{eq:matching_function_def}, and if $\mu_{x0}$ and $\mu_{0y}$ satisfy the following system
\begin{equation}
    \label{eq: nonlinear_system}
    \begin{cases}
n_{x}=\mu_{x0}+\sum_{y\in \mathcal{Y}}M_{xy}(\mu_{x0},\mu_{0y}),~\forall x\in \mathcal{X} \\
m_{y}=\mu _{0y}+\sum_{x\in \mathcal{X}}M_{xy}(\mu_{x0},\mu_{0y}),~\forall y\in \mathcal{Y}.
\end{cases}
\end{equation}
\end{definition}

Definition \ref{def: AMF} stresses that in the logit case, finding the aggregate equilibrium is equivalent to solving the system of nonlinear equations (\ref{eq: nonlinear_system})---a system of $\vert \mathcal{X}\vert
+\vert \mathcal{Y}\vert $ equations in the same number of
unknowns.

While existence and uniqueness of a solution to this problem follows from theorem~\ref{thm:existenceUniqueness}, it is instructive to give an alternative proof of the existence part, which is constructive and provides an algorithm to compute the equilibrium. The algorithm is described below:

\begin{algorithm}\label{algo:ippf}
\ \newline
\label{algorithma}
\begin{tabular}{l|p{5in}}
\raggedleft Step $0$ & Fix the initial value of $\mu_{0y} $ at $\mu_{0y}^0 = m_y$. \\
\raggedleft Step $t$
($t\geq 1$) & Keep the values $\mu_{0y}^{t-1}$ fixed. For each $x\in
\mathcal{X} $, solve for the value $\mu_{x0}^t$ of $\mu_{x0}$ such
that the equality $\mu_{x0}+\sum_{y\in \mathcal{Y}}M_{xy}(\mu_{x0},\mu_{0y}) =n_{x}$ holds. \\
 & Keep the values $\mu_{x0}^t$ fixed. For each $y\in
\mathcal{Y}$, solve for the value $\mu_{0y}^t$ of $\mu_{0y}$ such that the equality $\mu_{0y}+\sum_{x\in \mathcal{X}}M_{xy}(\mu_{x0},\mu_{0y}) =m_{y}$ holds.%
\end{tabular}
\ \newline
The algorithm terminates when $\sup_{y}|\mu^{t}_{0y} - \mu^{t-1}_{0y}|<\epsilon$.
\end{algorithm}

Algorithm~\ref{algo:ippf} converges to the unique solution to system (\ref{eq: nonlinear_system}), a result that we formally state in the following theorem.

\begin{theorem}\label{thm:mfe-existence}
Under assumptions \ref{ass:prefsheterog}  and \ref%
{ass:bargainingsets} and logit random utilities, algorithm \ref{algo:ippf} converges to a solution $(\mu_{x0}^\star,\mu_{0y}^\star)$ to the system of equations (\ref{eq: nonlinear_system}).
\end{theorem}

The proof of this result is straightforward after noticing that the $\mu^t_{0y}$ form a decreasing sequence bounded below by $0$. This implies that this sequence converges, which implies in turn that the sequence $\mu^t_{x0}$ also converges. Continuity of the matching functions then imply that the limits of these two sets of sequences are solution to system~\eqref{eq: nonlinear_system}.

In the transferable utility case, the two steps in algorithm~\ref{algo:ippf} can be expressed in closed form; this algorithm, initially described by~\textcite{galichonsalanie2022}, is an extension of the Iterated Proportional Fitting Procedure (IPFP), also known as Sinkhorn's algorithm, popular in the optimal transport literature.

Algorithm~\ref{algo:ippf} bears an interesting relation with the deferred acceptance algorithm of \textcite{galeshapley1962}, where, if the $x$'s form the proposing side, is such that initially, all the $y$'s are unassigned; like in the Gale and Shapley algorithm, the welfare of the $x$'s keep decreasing along the algorithm, while the welfare of the $y$'s keep increasing. Therefore, one may view algorithm~\ref{algo:ippf} as a continuous counterpart of the deferred acceptance algorithm.

\section{Estimation}

\label{sec:estimation} In this section, we follow a parametric approach and
assume that $(D_{xy})$ belongs to a parametric family $(D_{xy}^{\theta })$
where $\theta \in \mathbb{R}^{d}$ are the parameters of interest. We propose
to estimate these parameters by maximum likelihood.  Typically, empirical
applications will build on a model that specifies the surplus or payoff
functions of the agents on the market and mutually relates them. We gather
in the vector $\lambda \in \mathbb{R}^{K}$ the model's parameters that are
of interest to the econometrician. For instance, in the TU model, $\lambda $
may be a vector parameter such that the joint surplus has a linear
expression with respect to $\lambda $, so that $\Phi _{xy}=\sum_{k}\lambda_k \phi _{kxy}$; in the NTU\ specification, $\lambda $ can enter in the
specification of $\alpha $ and $\gamma $; etc. 

For a given value of $\lambda $, we can construct the bargaining sets $%
\mathcal{F}_{xy}^{\lambda }$ and compute the associated distance-to-frontier
functions $(\lambda ,u,v)\mapsto (D_{xy}(\lambda ,u_{x},v_{y}))_{xy}$. This,
in turn, allows us to compute the matching function, according to equation %
\eqref{matchingFunction}. In the remainder of this section, we will assume
sufficient smoothness on the parametrization, i.e. for each $xy\in \mathcal{%
XY}$, the map $(\lambda ,u,v)\mapsto (D_{xy}(\lambda ,u_{x},v_{y}))_{xy}$ is
twice continuously differentiable\footnote{This assumption is only really needed for estimation and inference purposes. Please note that this assumption might close the door to specific models (like NTU), but nothing more.} from $\mathbb{R}^{K}\times \mathbb{R}^{|%
\mathcal{X}|}\times \mathbb{R}^{|\mathcal{Y}|}$ to $\mathbb{R}^{|\mathcal{X}%
|\times |\mathcal{Y}|}$. 

We assume that we have a sample of $\hat{N}$ i.i.d.~draws of household types 
$xy\in \mathcal{XY}\cup \mathcal{X}_{0}\cup \mathcal{Y}_{0}$, so that we
observe the matching $\hat{\mu}=\{\hat{\mu}_{xy}\}_{xy\in \mathcal{XY}},\{%
\hat{\mu}_{x0}\}_{x\in \mathcal{X}},\{\hat{\mu}_{0y}\}_{y\in \mathcal{Y}}$.
We can introduce%
\begin{equation*}
\hat{\pi}_{xy}=\frac{\hat{\mu}_{xy}}{\hat{N}},~\hat{\pi}_{x0}=\frac{\hat{\mu}%
_{x0}}{\hat{N}}\text{ and }\hat{\pi}_{0y}=\frac{\hat{\mu}_{0y}}{\hat{N}}
\end{equation*}%
the empirical frequencies in the sample. Given the parameter value $(\lambda
,u,v)$, the total predicted number of households will be 
\begin{equation*}
N(\lambda ,u,v)=\sum_{xy\in \mathcal{XY}}e^{-D_{xy}(\lambda
,u,v)}+\sum_{x\in \mathcal{X}}e^{-u_{x}}+\sum_{y\in \mathcal{Y}}e^{-v_{y}},
\end{equation*}%
and the probability of observing household $xy$ (respectively $x0$ and $0y$)
is given by 
\begin{equation*}
\pi _{xy}(\lambda ,u,v)=\frac{e^{-D_{xy}(\lambda ,u,v)}}{N(\lambda ,u,v)},%
\text{ }\pi _{x0}(\lambda ,u,v)=\frac{e^{-u_{x}}}{N(\lambda ,u,v)}\text{ and 
}\pi _{0y}(\lambda ,u,v)=\frac{e^{-v_{y}}}{N(\lambda ,u,v)},
\end{equation*}%
so, assuming that the observations were sampled in an i.i.d. fashion from
distribution $\pi $ at the true value of the parameter, the log-likelihood
of the sample is (after a rescaling by $\hat{N}$)%
\begin{equation*}
l(\lambda ,u,v)=-(\sum_{xy\in \mathcal{XY}}\hat{\pi}_{xy}D_{xy}(\lambda
,u,v)+\sum_{x\in \mathcal{X}}\hat{\pi}_{x0}u_{x}+\sum_{y\in \mathcal{Y}}\hat{%
\pi}_{0y}v_{y}+\log N(\lambda ,u,v)).
\end{equation*}

\bigskip 

We can now investigate the properties of the maximum likelihood. Letting $%
\theta =(\lambda ,u,v)$, the first order conditions with respect to $u_{x}$, 
$v_{y}$ and $\lambda _{k}$ yield respectively%
\begin{equation}
\left\{ 
\begin{array}{l}
\sum_{xy\in \mathcal{XY}\cup \mathcal{X}_{0}\cup \mathcal{Y}_{0}}\hat{\pi}%
_{xy}\frac{\partial D_{xy}(\theta )}{\partial \lambda _{k}}=\sum_{xy\in 
\mathcal{XY}\cup \mathcal{X}_{0}\cup \mathcal{Y}_{0}}\pi _{xy}(\theta )\frac{%
\partial D_{xy}(\theta )}{\partial \lambda _{k}} \\ 
\sum_{xy\in \mathcal{XY}}\hat{\pi}_{xy}\frac{\partial D_{xy}(\theta )}{%
\partial u_{x}}+\hat{\pi}_{x0}=\sum_{xy\in \mathcal{XY}}\pi _{xy}(\theta )%
\frac{\partial D_{xy}(\theta )}{\partial u_{x}}+\pi _{x0}(\theta ) \\ 
\sum_{xy\in \mathcal{XY}}\hat{\pi}_{xy}\frac{\partial D_{xy}(\theta )}{%
\partial v_{y}}+\hat{\pi}_{0y}=\sum_{xy\in \mathcal{XY}}\pi _{xy}(\theta )%
\frac{\partial D_{xy}(\theta )}{\partial v_{y}}+\pi _{0y}(\theta ).%
\end{array}%
\right.   \label{foc-mle}
\end{equation}

The score of observation $a=xy\in \mathcal{XY}\cup \mathcal{X}_{0}\cup 
\mathcal{Y}_{0}$ is given by $\nabla _{\theta }\ln \pi _{a}(\theta )$, that
is%
\begin{eqnarray*}
\frac{\partial \ln \pi _{a}(\theta )}{\partial \lambda _{k}} &=&-\frac{%
\partial D_{a}(\theta )}{\partial \lambda _{k}}+\mathbb{E}_{\pi (\lambda
,u,v)}\left[ \frac{\partial D_{\tilde{a}}}{\partial \lambda _{k}}(\lambda
,u,v)\right]  \\
\frac{\partial \ln \pi _{a}(\theta )}{\partial u_{x}} &=&-\frac{\partial
D_{a}}{\partial u_{x}}(\lambda ,u,v)+\mathbb{E}_{\pi (\lambda ,u,v)}\left[ 
\frac{\partial D_{\tilde{a}}}{\partial u_{x}}(\lambda ,u,v)\right]  \\
\frac{\partial \ln \pi _{a}(\theta )}{\partial v_{y}} &=&-\frac{\partial
D_{a}}{\partial v_{y}}(\lambda ,u,v)+\mathbb{E}_{\pi (\lambda ,u,v)}\left[ 
\frac{\partial D_{\tilde{a}}}{\partial v_{y}}(\lambda ,u,v)\right] 
\end{eqnarray*}%
where in expression $\mathbb{E}_{\pi (\lambda ,u,v)}\left[ f_{\tilde{a}%
}(\lambda ,u,v)\right] $, $\tilde{a}$ is seen as a random variable drawn
from set $\mathcal{XY}\cup \mathcal{X}_{0}\cup \mathcal{Y}_{0}$ using
probability $\pi (\lambda ,u,v)$, so that the expectation means $\sum_{a\in 
\mathcal{XY}\cup \mathcal{X}_{0}\cup \mathcal{Y}_{0}}\pi _{a}(\lambda
,u,v)f_{a}(\lambda ,u,v)$. As a result, the Fisher information matrix $%
\mathcal{I}(\lambda ,u,v)$ is given by%
\begin{equation*}
\mathcal{I}(\theta )=\mathbb{E}_{\pi (\lambda ,u,v)}[\nabla _{\theta }\ln
\pi _{\tilde{a}}(\theta )(\nabla _{\theta }\ln \pi _{\tilde{a}}(\theta
))^{\top }].
\end{equation*}

\bigskip 

We recapitulate the previous results as follows

\begin{theorem}
\label{thm:MLE}\phantom{ }

(i) The log-likelihood is expressed using%
\begin{equation}
l(\theta )=-(\sum_{xy\in \mathcal{XY}}\hat{\pi}_{xy}D_{xy}(\theta
)+\sum_{x\in \mathcal{X}}\hat{\pi}_{x0}u_{x}+\sum_{y\in \mathcal{Y}}\hat{\pi}%
_{0y}v_{y}+\log N(\theta )),  \label{ExprLogLikelihood2}
\end{equation}%
which has first order conditions~(\ref{foc-mle}).

(ii) Letting $\hat{\theta}$ be the maximum likelihood estimator, $N^{1/2}(%
\hat{\theta}-\theta )\Rightarrow \mathcal{N}(0,V_{\theta })$ as the sample
size $\hat{N}\rightarrow +\infty $, where the variance-covariance matrix $%
V_{\theta }$ can be consistently estimated by%
\begin{equation}
\hat{V}_{\theta }=\mathcal{I}(\hat{\theta})^{-1}.  \label{asymptoticVar}
\end{equation}
\end{theorem}
\section{Remarks and discussions}

\subsection{Positive assortative matching}
Positive assortative matching (PAM) is well studied in the TU case. There, supermodularity of the surplus function ensures positive assortative matching. In the context of imperfectly transferable utility models, assortativeness has been studied in e.g. \textcite{legrosnewman2007}, \textcite{chiapporireny2016}, \textcite{chadeeeckhout2017} and \textcite{chiappori2017}.  We now explore some of these ideas, following a similar approach to that of \textcite[][chapter 7]{chiappori2017}. For simplicity, we will assume that men and women differ along a single dimension $x\in\mathbb{R}$ and $y\in\mathbb{R}$, respectively. Let $V(x,u(x),y)$ denote the maximum utility a woman $y$ can receive when matched with man $x$ set to receive utility $u(x)$, and let $F(x,y)=V(x,u(x),y)$.
A woman $y$ chooses her preferred man and receives payoff
\begin{equation*}
    v\left( y\right) =\max_{x}F\left( x,y\right)
\end{equation*}
which yields the first order condition
\begin{equation*}
    \partial _{x}F\left( x,T\left( x\right) \right) =0
\end{equation*}
where $T(.)$ is the assignment rule. Thus $\partial
_{x}V\left( x,u\left( x\right) ,y\right) +\partial _{u}V\left( x,u\left(
x\right) ,y\right) u^{\prime }\left( x\right) =0$, hence $u^{\prime }\left( x\right) =-\frac{\partial _{x}V\left( x,u\left( x\right)
,y\right) }{\partial _{u}V\left( x,u\left( x\right) ,y\right) }$. Since the first order condition holds for all $x$, then differentiating a second time with respect to $x$ yields $\partial _{xx}^{2}F\left( x,T\left( x\right) \right) +\partial
_{xy}^{2}F\left( x,T\left( x\right) \right) T^{\prime }\left( x\right) =0$, that is
\begin{equation*}
    T^{\prime }\left( x\right) =-\frac{\partial _{xx}^{2}F\left( x,T\left(
x\right) \right) }{\partial _{xy}^{2}F\left( x,T\left( x\right) \right) }
\end{equation*}
From the second order condition, $\partial _{xx}^{2}F\left( x,T\left( x\right) \right) \leq 0$, thus $T^{\prime }\left( x\right)$ is of the same sign as $\partial _{xy}^{2}F\left( x,T\left( x\right) \right)$, that is, we have PAM whenever
\begin{equation}\label{eq:pam}
\partial _{xy}^{2}V\left( x,u\left( x\right) ,y\right) - \partial
_{uy}V\left( x,u\left( x\right) ,y\right) \frac{\partial _{x}V\left(
x,u\left( x\right) ,y\right) }{\partial _{u}V\left( x,u\left( x\right)
,y\right) }\geq 0
\end{equation}

As pointed out in \textcite{chiappori2017}, this restriction is a generalization of the famous Spence-Mirrlees condition. In particular, in the TU case when $V(x,u(x),y) = \Phi(x,y)-u(x)$ (where $\Phi(x,y)$ denotes the marital surplus), we have $\partial
_{uy}V\left( x,u\left( x\right) ,y\right) = 0$ and equation \eqref{eq:pam} boils down to the usual supermodularity condition on the surplus function.
In addition, note that having $\partial _{xy}^{2}V\left( x,u\left( x\right) ,y\right) \geq 0$ (i.e. women prefer high-type men, even more so when they are high-type themselves), $\partial _{x}V\left(
x,u\left( x\right) ,y\right) \geq 0$ and $\partial
_{uy}V\left( x,u\left( x\right) ,y\right)\geq 0$ (i.e. it is easier for high-type women to bid for men as the frontier is flatter) would result in PAM, unsurprisingly.

\subsection{Aggregate matching equilibrium beyond the logit case}

The aggregate matching equation formulation consists in rewriting the 
the system of equations in definition~\ref{def:equil} as a simpler system of equations that only depends on the matching $%
\mu $. To achieve this, we introduce the
Legendre-Fenchel transform (a.k.a.~convex conjugate) of $G$ and $H$:
\begin{equation}
G^{\ast }( \mu ) =\sup_{U}\left\{ \sum_{xy}\mu
_{xy}U_{xy}-G( U) \right\} \quad \text{and}\quad H^{\ast }(
\nu ) =\sup_{V}\left\{ \sum_{xy}\nu _{xy}V_{xy}-H( V)
\right\} .  \label{GstarandHstar}
\end{equation}%
Under smoothness assumptions (that hold here given assumption~\ref%
{ass:Distrib}), results from convex analysis \parencite[see][]{rockafellar1970} indicate that
\begin{equation*}
\mu =\nabla G(U)\iff U=\nabla G^{\ast }(\mu )\text{ and }\nu =\nabla
H(V)\iff V=\nabla H^{\ast }(\nu ).
\end{equation*}%
Using these results, we can express $U$ and $V $ as functions of $\mu $ and substitute these out in the system of equations in definition~\ref{def:equil}. 
Equilibrium can then be characterized by a set of $\left\vert \mathcal{X}\right\vert \times
\left\vert \mathcal{Y}\right\vert $ equations that only depends on $%
\mu $. GKW conclude that a matching $\mu \in \mathcal{M%
}^{\interior}$ is an aggregate equilibrium matching if and only if
\begin{equation}
D_{xy}\left( \frac{\partial G^{\ast }( \mu ) }{\partial \mu _{xy}}%
,\frac{\partial H^{\ast }( \mu ) }{\partial \mu _{xy}}\right) =0%
\text{ for all }x\in \mathcal{X},~y\in \mathcal{Y}.  \label{master}
\end{equation}
(see GKW, proposition 1).

\subsection{Comparative statics}
In this section we examine how changes in exogenous parameters affect
the matching numbers $\mu _{xy}$ and the equilibrium utilities $U_{xy}$ and $V_{xy}$. The framework can be used to obtain comparative statics results that are relevant in many applications, e.g. to study the impact of a change in the distance function or in the parameters of the underlying model (for example, a change in the quantities $\alpha$ and $\gamma$ in the examples of section \ref{sec:examples}). For the sake of brevity, we will only study the effect of changes in the distributions of characteristics of the populations. Building on the equilibrium formulation provided in the previous section, we provide comparative statics
results, which allows us to predict the vector of change in $\delta \mu _{xy}$ in the number of matched pairs at equilibrium as a function of the change in the number of men and women of each types, $\delta n_{x}$
and $\delta m_{y}$. From the expression of $\delta \mu $, we can recover the expression of the
systematic utilities at equilibrium $\delta U$ and $\delta V$. The following results extend those of \textcite{galichonsalanie2022} to the case with imperfectly transferable utility.

In order to
do this, we shall use the \emph{vectorized} elements $\mu$, $U$ and $V$ and use the following notations. First, $\delta \mu $, $\delta U$, and $\delta V$ are the doubly-indexed vectors whose $\left( xy\right) $-th entry
are respectively $\delta \mu _{xy}$, $\delta U_{xy}$, and $\delta V_{xy}$. Then, we let $\partial _{u}D $ (resp.~$\partial _{v}D $) be the
doubly-indexed matrix whose $\left( xy\right) \left( x^{\prime }y^{\prime
}\right) $-th entry is $\partial _{u}D _{xy}$ (resp. $\partial _{v}D _{xy}$) if $x=x^{\prime }$ and $%
y=y^{\prime }$, $0$ otherwise. We denote $\partial^{2}G^{\ast }$ (resp.~$\partial^{2}H^{\ast }$) the doubly-indexed matrix
whose $\left( xy\right) \left( x^{\prime }y^{\prime }\right) $-th entry is $%
\partial ^{2}G^{\ast }\left( \mu \right) /\partial \mu _{xy}\partial \mu
_{x^{\prime }y^{\prime }}$ (resp.~$\partial ^{2}H^{\ast }\left( \mu \right)
/\partial \mu _{xy}\partial \mu _{x^{\prime }y^{\prime }}$). Finally, we let $\frac{\mu \delta n}{n}$ (resp.~$\frac{\mu \delta m}{m}$) be the
doubly-indexed vector whose $\left( xy\right) $-th term is $\mu _{xy}\delta
n_{x}/n_{x}$ (resp.~$\mu _{xy}\delta m_{y}/m_{y}$).

Using these notations, we now focus on the effect of a change in the population on the equilibrium outcome. As \textcite[][p.~121]{becker1993} puts it:

\begin{quotation}
``[\ldots] An increase in the number of men of a particular quality tends to
lower the incomes of all men and raise those of all women because of the
competition in the marriage market between men and women of different
qualities.''\footnote{``Incomes'' is to be understood as ``utilities'' in the context of this citation.}
\end{quotation}

We formalize this Beckerian intuition in the following result.

\begin{theorem}
\label{thm:compstats}Suppose assumptions \ref{ass:prefsheterog}, \ref%
{ass:Distrib}, and \ref{ass:bargainingsets} hold and let $\left( \mu ,U,V\right) $ be the unique equilibrium outcome. Assume that $n$ and $m$ are changed by some infinitesimal quantities $\delta n$ and $\delta m$, respectively. Then:

(i) The change in $\mu $ is given by 
\begin{equation}\label{deltaMu}
\delta \mu = \left( \partial _{u}D \partial^{2}G^{\ast }+\partial _{v}D
\partial^{2}H^{\ast }\right) ^{-1}   \cdot 
\left[ \partial _{u}D \partial^{2}G^{\ast }\frac{%
\mu \delta n}{n} +\partial _{v}D \partial^{2}H^{\ast }\frac{\mu \delta m}{m}\right]
\end{equation}

(ii) The changes in $U$ and $V$ are given as a function of $\delta \mu $, $\delta n$ and 
$\delta m$ by 
\begin{equation}
\delta U=\left( \partial^{2}G^{\ast }\right) \left( \delta \mu -\frac{\mu \delta n}{%
n}\right) \text{ and }\delta V=\left( \partial^{2}H^{\ast }\right) \left( \delta
\mu -\frac{\mu \delta m}{m}\right) .  \label{deltaUV}
\end{equation}

(iii) In particular, under transferable utility, equation \eqref{deltaMu} becomes%
\begin{equation*}
\delta \mu =\left( \partial^{2}G^{\ast }+\partial^{2}H^{\ast }\right) ^{-1}\left[
\partial^{2}G^{\ast }\frac{\mu \delta n}{n}+\partial^{2}H^{\ast }\frac{\mu \delta m}{m}%
\right] .
\end{equation*}
\end{theorem}
To clarify the structure of our comparative statics, we look at an example
with logit unobserved heterogeneity and no observable heterogeneity. 

\begin{example}[Comparative statics]
\label{ex:averse1} Consider an ITU-logit model and assume that there is only one type of man and one type of
woman. Let $\mu$ be the number of married couples. In this case, the
systematic part of the equilibrium utilities of married men and women are $%
U=\log \mu /\left( n-\mu \right) $ and $V=\log \mu /\left( m-\mu \right) $,
and $\partial^{2}G^{\ast }=\frac{n}{\mu \left( n-\mu \right) }$ and $\partial^{2}H^{\ast }=%
\frac{m}{\mu \left( m-\mu \right) }$. Then, equation \eqref{deltaMu}
simplifies to%
\begin{equation*}
\delta \mu =\mu \frac{\partial _{u}D \left( m-\mu \right) \delta
n+\partial _{v}D \left( n-\mu \right) \delta m}{\partial _{u}D \left(
m-\mu \right) n+\partial _{v}D \left( n-\mu \right) m}.
\end{equation*}%
We have 
\begin{equation*}
\frac{\delta \mu }{\mu }=\theta \frac{\delta n}{n}+\left( 1-\theta \right) 
\frac{\delta m}{m},
\end{equation*}%
where we take 
\begin{equation*}
\theta \equiv \frac{\partial _{u}D \left( m-\mu \right) n}{\partial
_{u}D \left( m-\mu \right) n+\partial _{v}D \left( n-\mu \right) m}.
\label{theta}
\end{equation*}%
Thus, we can compute that 
\begin{equation*}
\delta U=\frac{n\left( 1-\theta \right) }{n-\mu }\left( \frac{\delta m}{m}-%
\frac{\delta n}{n}\right) \text{ and }\delta V=\frac{m\theta }{m-\mu }%
\left( \frac{\delta n}{n}-\frac{\delta m}{m}\right) ,
\end{equation*}%
from which we see that if $\frac{\delta m}{m}>\frac{\delta n}{n}$ (i.e., if
the women's relative increase in population is larger than the men's), then
the systematic utility of men increases, and the systematic utility of women
decreases. (The converse is also true.)
\end{example}

Finally, we consider the effect of the change in population supplies on the average welfare of a man of type $x$ and a woman of type $y$, defined as
\begin{equation*}
u_{x}=G_{x}\left( U\right) \text{ and }v_{y}=H_{y}\left( V\right)
\end{equation*}%
We obtain the following result:

\begin{corollary}[Unexpected Symmetry or Expected Dissymmetry]
\label{cor:unexpectedSymmetry}Maintain assumptions \ref{ass:prefsheterog}, \ref{ass:Distrib}, and \ref{ass:bargainingsets}. Assume that the population numbers vary by $%
\delta n$ and $\delta m$. Then the ``unexpected symmetry'' result of \textcite{deckerliebmccannetal2013} holds within men and within women, namely%
\begin{equation*}
\frac{\partial u_{x}}{\partial n_{x^{\prime }}}=\frac{\partial u_{x^{\prime
}}}{\partial n_{x}}\text{ and }\frac{\partial v_{y}}{\partial m_{y^{\prime }}%
}=\frac{\partial v_{y^{\prime }}}{\partial m_{y}};
\end{equation*}
however, symmetry does not necessarily hold \emph{across} men and women, i.e.%
\begin{equation*}
\frac{\partial u_{x}}{\partial m_{y}}\text{ does not always coincide with }%
\frac{\partial v_{y}}{\partial n_{x}}\text{,}
\end{equation*}%
even though it does in the TU case.
\end{corollary}

Note that corollary \ref{cor:unexpectedSymmetry} provides testable implications for TU. Assuming the econometrician can observe exogenous changes in populations and measure average welfare, it is in principle possible to test for symmetry, which holds across men and women under TU but does necessarily hold under general ITU.

\subsection{Equilibrium vs optimality}

One of the many appealing properties of TU is the fact that equilibrium and optimality coincide. We shall discuss this statement in greater detail, both in the case with and without heterogeneity, and contrast with the ITU case.

In the case without heterogeneity, it is well known that the TU matching model can be reformulated as an optimal assignment model (or problem), in which a social planner maximizes total surplus under feasibility conditions on the matching. The optimal assignment problem is written as a linear programming problem, in which the 
complementary slackness conditions for optimality coincide with our equilibrium conditions. This is specific to the TU case, however ; in the ITU case, equilibrium conditions are \emph{%
not}, in general, the first-order conditions associated to an optimization problem. That is, equilibrium cannot be interpreted as the solution to some welfare maximization problem.

In the case with unobserved heterogeneity, things are no different. Recall that a matching $\mu \in \mathcal{M%
}^{\interior}$ is an aggregate equilibrium matching if and only if
\begin{equation}
D_{xy}\left( \frac{\partial G^{\ast }( \mu ) }{\partial \mu _{xy}}%
,\frac{\partial H^{\ast }( \mu ) }{\partial \mu _{xy}}\right) =0%
\text{ for all }x\in \mathcal{X},~y\in \mathcal{Y}.  \label{masterbis}
\end{equation}
which, in the TU case, boils down to
\begin{equation}
   \nabla G^{\ast
}( \mu ) +\nabla H^{\ast }( \mu ) =\Phi _{xy} 
\label{eq:mastertu}
\end{equation}
It turns out that equation \eqref{eq:mastertu} coincides with the first order conditions associated with the following welfare maximization problem
\begin{equation*}
\max_{\mu }\left\{ \sum_{xy}\mu _{xy}\Phi _{xy}-\mathcal{E}\left( \mu
\right) \right\} ,
\end{equation*}%
where $\mathcal{E}:=G^{\ast }+H^{\ast }$ is an entropy penalization term. This is again a specific feature of the TU model. In general, equation \eqref{masterbis} cannot be interpreted as the first-order conditions of an optimization problem because the Jacobian of the function defined by the left hand-side
of equation \eqref{masterbis} is not symmetric.

\subsection{Case without unobserved heterogeneity}
\label{sec:unregularizedcase}
In the case when there is no heterogeneity, but when individuals are clustered into types $x\in \mathcal X$ for the men and  $y\in \mathcal Y$ for the women, definition~\ref{def:ITU-none} has a straightforward extension which is summarized into table~\ref{table:eq-matching} below. An \emph{equilibrium outcome} is a vector $(\mu,u,v)$ solution to this set of equations.
\begin{table}[H]
\centering
\begin{tabular}{ | l | l | l |}
\hline
\textbf{Variable} & \textbf{Constraint} & \textbf{Complementarity} \\ 
\hline
$\mu _{xy}\geq 0$  & $D_{xy}\left( u_{x},v_{y}\right) \geq 0$  & $%
\sum_{xy}\mu _{xy}D_{xy}\left( u_{x},v_{y}\right) =0$ \\ 
\hline
$\mu _{x0}\geq 0$  & $u_{x}\geq 0$ & $\sum_{x\in \mathcal{X}}\mu _{x0}u_{x}=0
$ \\ 
\hline
$\mu _{0y}\geq 0$  & $v_{y}\geq 0$ & $\sum_{y\in \mathcal{Y}}\mu _{0y}v_{y}=0
$ \\ 
\hline
$u_{x}$ & $\sum_{y}\mu _{xy}+\mu _{x0}=n_{x}$ &  NA \\ 
\hline
$v_{y}$ & $\sum_{y}\mu _{xy}+\mu _{x0}=m_{y}$ & NA \\
\hline
\end{tabular}

\caption{Equilibrium matching with imperfectly transferable utility}\label{table:eq-matching}
\end{table}
As one can see on table~\ref{table:eq-matching}, the first column contains the variables (``primal'' or quantity-like $\mu$, as well as ``dual'' or price-like $u$ and $v$), the second contains the corresponding constraints, and the third column  represents the complementary slackness condition. In the TU case, these relations are simply the optimality conditions associated with a linear programming problem; in the general ITU case, one loses the reference to optimization.

\bigskip 

Several approaches exist to prove that an equilibrium outcome $\left( \mu
,u,v\right) $ exists. When $n_{x}=1$ and $m_{y}=1$ for all $x$ and $y$,
\textcite{kelsocrawford1982}'s approach consists in discretizing the set of $v_{y}$
into a finite set of possible values $v_{y}^{1}=0<v_{y}^{2}<...<v_{y}^{K}$
and consider a deferred acceptance algorithm where the $x$'s propose. In an
initial round, the $v_{y}$'s are set to their smallest value $v_{y}^{1}=0$.
Each $x$ then makes an offer to the $y$ maximizing $u_{xy}$ where $%
u_{xy}^{k}$ is such that $D_{xy}\left( u_{xy}^{k},v_{y}^{k}\right) $. If no $%
y$ received more than one offer, the algorithm stops. If some $y$'s are made
more than one offer, then the values of $v_{y}$ for these $y$ go up by one
step, and the proposal phase is repeated. The algorithm runs in that way
until each $y$ receives at most one offer. \textcite{hatfieldmilgrom2005} provide a reformulation of Kelso and Crawford's algorithm in lattice-theoretic terms. 

\textcite{scarf1967} using a combinatorial lemma which also provides a computational
argument. The idea behind Scarf's argument is to look for two basis of variables. First, a basis that guarantees primal feasibility, meaning that all the constraint that only involve $\mu$ in table~\ref{table:eq-matching}  are satisfied; such a problem alone is not difficult. Second, a basis that guarantees dual feasibility, which deals with all the constraints in table~\ref{table:eq-matching} involving variables $u$ and $v$ only; again, this problem alone is not hard. As shown by Scarf, the problem of complementarity consists of ensuring that these two basis coincide. Scarf shows that it is quite straightforward to construct a primal feasible basis and a dual feasible basis that have all but one element in common. He then proceeds to show that a pivoting algorithm in the spirit of \textcite{lemkehowson1964} algorithm initialized on these overlapping basis necessarily terminates when the two basis coincide.

In \textcite{galichonkominersweber2015}, it is shown that the proof of existence of a solution to the unregularized problem described in table \ref{table:eq-matching} can be deduced very simply from
the existence of a solution to the regularized problem of section \ref{sec:aggreq}. Indeed, one can show that if we take as function $M_{xy}(\mu_{x0},\mu_{0y}) = \exp ( -D_{xy}(-\log \mu_{x0}, -\log \mu_{0y} ) / \sigma ) $, which recovers \eqref{matchingFunction} when $\sigma = 1$, a matching function equilibrium exists for any value of $\sigma >0$, and when $\sigma \to 0$, the solution to the matching function equilibrium problem tends to a solution to the unregularized equilibrium matching problem.

Other approaches to the existence of an equilibrium matching outcome exist.  
\textcite{alkan1989} studies existence in this problem using a continuation method
starting from the TU case.  \textcite{alkangale1990} use a non-constructive topological argument.

\subsection{Normative properties of equilibrium}
\label{par:normative-properties}
In this section we discuss how our notion of equilibrium fares with respect to two important normative properties, namely equal treatment and Pareto efficiency.

\subsubsection{Equal treatment}
The property of equal treatment means that any two individuals on the market
who are perfectly similar receive the same payoff at equilibrium. This
property is often quoted as a desirable feature of a centralized mechanism,
because it conveys a sense of fairness, or at the least, absence of
arbitrariness. However, the focus of our chapter is not centralized
mechanisms, but decentralized market equilibria. Here, we discuss how the equal treatment property
arises in the type of equilibria we study.

1. First, when all the distance functions $(u,v)\rightarrow D_{ij}\left(
u,v\right) $ are strictly increasing both in $u$ and in $v$ (i.e. when the
frontier of the feasible set is downward-sloping and does not feature vertical or horizontal segments, which means that it is not
possible to increasing one partner's utility without decreasing the other
one's), then equal treatment holds. Indeed, say two individuals $i$ and $%
i^{\prime }$ in $\mathcal{I}$ are such that $D_{ij}\left( .,.\right)
=D_{i^{\prime }j}\left( .,.\right) $ for all $j\in \mathcal{J}$, meaning
that $i$ and $i^{\prime }$ are perfect substitutes from the point of view of
the other side of the market. If $u_{i}$ and $u_{i^{\prime }}$ were to
differ, say $u_{i}>u_{i^{\prime }}$ this would imply (as $u_{i^{\prime
}}\geq \mathcal{U}_{i^{\prime }0}$) that $i$ is matched with some $j\in 
\mathcal{J}$. We have $0=D_{ij}\left( u_{i},v_{j}\right) =D_{i^{\prime
}j}\left( u_{i},v_{j}\right) $ because $i$ and $j$ are matched and because $i
$ and $i^{\prime }$ are perfect substitutes. But we have $D_{i^{\prime
}j}\left( u_{i},v_{j}\right) >D_{i^{\prime }j}\left( u_{i}^{\prime
},v_{j}\right) \geq 0$ where the first inequality comes from $%
u_{i}>u_{i^{\prime }}$, and the second one from the stability requirement,
which brings a contradiction. As a result, if two individuals are perfectly
similar, then their equilibrium payoff should be the same, which is the very
definition of equal treatment.

2a. Things are, however, slightly more subtle when the frontier of the
feasible set can accommodate horizontal or vertical parts, which means that
it is possible to bring changes to some agents' welfare without affecting their
partner's welfare. This arises, for instance, in the case of
non-transferable utility discussed in section \ref{ex:NTU}, which is to say, when $%
D_{ij}\left( u,v\right) =\max \left\{ u_{i}-\alpha _{ij},v_{j}-\gamma
_{ij}\right\} $. In that case, there may exist two identical agents $i$ and $%
i^{\prime }$ such that $u_{i}\neq u_{i^{\prime }}$ in an equilibrium
matching. For instance, assume $\mathcal{I}=\left\{ 1,2\right\} $ and $%
\mathcal{J}=\left\{ 1\right\} $ and $\alpha _{11}=\alpha _{21}=1$ and $%
\gamma _{11}=\gamma _{21}=1$, and the reservation utilities are set to zero,
then $\mu _{ij}=1$ if $i=1$ and $j=1$, $0$ otherwise and $u_{1}=1,u_{2}=0$
and $v_{1}=1$ is an equilibrium matching, but does not verify equal
treatment. However, it is worth noting that there is one solution that
verifies equal treatment: indeed, $\mu _{ij}=1$ if $i=1$ and $j=1$, $0$
otherwise and $u_{1}=0,u_{2}=0$ and $v_{1}=1$ is an equilibrium matching
which does satisfy equal treatment.

2b. Although there are solutions without equal treatment, we would like to argue
that the equilibrium with equal treatment is probably more satisfactory as a
modelling device, given that there will be a form of competition among
agents if similar agents are faced with different welfare outcomes, and this
competition will equalize ex-post utilities, whether it is carried through
waiting lines, physical fight, overinvestment, advertising, etc. This is the
reason we build equal treatment into our definition of an equilibrium
matching, and consider $x$ to be the equivalence class of the $i$'s such
that $D_{ij}\left( .,.\right) =D_{i^{\prime }j}\left( .,.\right) $ for all $%
j\in \mathcal{J}$, and consider $y$ to be the equivalence class of the $j$'s
such that $D_{ij}\left( .,.\right) =D_{ij^{\prime }}\left( .,.\right) $ for
all $i\in \mathcal{I}$. In that case, we denote $n_{x}$ the number of $i$'s
such that $x_{i}=x$ and $m_{y}$ the number of $j$'s such that $y_{j}=y$, and
we can request $u_{i}$ and $v_{j}$ to depend respectively on $x_{i}=x$ and
on $y_{j}=y$ only. One can then define an equilibrium matching as an outcome 
$\left( \mu ,u,v\right) $ that satisfies 
\begin{eqnarray*}
&&\mu \in \mathcal{M}\left( n,m\right)  \\
&&u_{x}\geq \mathcal{U}_{x0}\text{ and }v_{y}\geq \mathcal{V}_{0y} \\
&&D_{xy}\left( u_{x},v_{y}\right) \geq 0 \\
&&\mu _{xy}>0\implies D_{xy}\left( u_{x},v_{y}\right) =0
\end{eqnarray*}%
and as discussed in section \ref{sec:existenceuniqueness}, one can show that this problem has a
solution under the assumptions made on $D_{xy}$ in the chapter.

To summarize, under the assumptions made in the chapter there is an
equilibrium matching which satisfies equal treatment, and under somewhat
stronger assumptions (namely, that the utility frontiers are strictly
downward sloping and do not feature vertical or horizontal segments), any equilibrium matching satisfies equal treatment.

\subsubsection{Pareto efficiency}
As is well-known, a  feasible outcome is Pareto efficient if there is no other feasible outcome which every agent on both sides of the market weakly prefer, while some prefer it strictly. An adaptation of the argument in the previous paragraph shows that if all the distance functions $(u,v) \to D_{ij}(u,v)$  are strictly increasing in both arguments, then any equilibrium outcome is Pareto efficient. Indeed, due to the assumption made, any increase in the utility of one agent would immediately result in a decrease in the utility of that agent's matched partner.

\subsection{One-to-many models}\label{sec:onetomany}

We can extend the technology developed in this chapter (distance functions and matching functions) in order to extend the setting to one-to-many models of matchings, which is particularly useful when considering models of the labor market. In the paper~\textcite{corblet2023} taken from her doctoral thesis, Pauline Corblet introduced a models of one-to-many matching with random utility and transferable utility. Here, we extend her insights to the case of imperfectly transferable utility. Assume that instead of hiring a single worker, a firm may employ a
bundle of workers. This bundle of workers is a vector $b\in \mathbb{N}^{%
\mathcal{X}}$, where, for each worker type $x\in \mathcal{X}$, $b_{x}$ is
the number of workers $x$ employed by the firm. In a transferable utility
setting, one would specify $\Phi _{by}$ which is the total amount of output
to be shared between the firm $y$ and the workers of the bundle $b$. The
frontier of the feasible set of utilities $(u,v)$ for a bundle $b$ is then
such that $\Phi _{by}=\sum_{x\in \mathcal{X}}b_{x}u_{x}+v_{y}$, and as a
result, if we extend definition~\ref{def:distance-to-frontier} to the one-to-many setting, the
distance-to-frontier function would then be%
\begin{equation}
D_{by}(u,v)=\frac{\sum_{x\in \mathcal{X}}b_{x}u_{x}+v_{y}-\Phi _{by}}{s_{b}+1%
},  \label{DTF-OTM}
\end{equation}%
where $s_{b}=\sum_{x}b_{x}$ is the size of the bundle, or the total number
of workers in bundle $b$. This extends in a straightfoward manner to the
imperfectly transferable utility case. Letting $D_{by}(u,v)$ be a function
which is nondecreasing in $v_{y}$ and in $u_{x}$ for any $x$ such that  $%
b_{x}>0$, which does not depend on $u_{x}$ whenever $b_{x}=0$, and imposing
that 
\begin{equation*}
D_{by}(u+t\mathbf{1}_{\mathcal{X}},v+t)=D_{by}(u,v)+t,
\end{equation*}%
the arguments in the chapter extend to that case and lead to a mass of
matches between workers bundle $b$ and firm $y$ to be $\mu _{by}$ given by  
\begin{equation*}
\left\{ 
\begin{array}{l}
\mu _{by}=\exp (-D_{by}(u,v)) \\ 
\mu _{x0}=\exp (-u_{x})%
\end{array}%
\right. 
\end{equation*}%
where $\mu _{x0}$ is the mass of unassigned workers of type $x$ (assuming
that their reservation utility is normalized to zero). The mass of void
firms (i.e. firms that do not hire any worker) is given by $\mu _{by}$ where 
$b=0^{\mathcal{X}}$. In that case, the market clearing conditions yield%
\begin{equation*}
\left\{ 
\begin{array}{l}
\sum_{y\in \mathcal{Y}}\sum_{b\in \mathbb{N}^{\mathcal{X}}}b_{x}\exp
(-D_{by}(u,v))+\exp (-u_{x})=n_{x} \\ 
\sum_{b\in \mathbb{N}^{\mathcal{X}}}\exp (-D_{by}(u,v))=m_{y}.%
\end{array}%
\right. 
\end{equation*}
The question of existence and uniqueness of solutions to such a system is more delicate than in the one-to-one case, due to the co-existence of complementarities between workers. It is beyond the scope of this chapter. 

\subsection{Full assignment}

In all the models examined so far, agents on the market had the option to remain single. We call these models matching models with partial assignment \parencite{chenchoogalichonetal2023b}. In this section, we discuss matching models with full assignment instead, i.e. models in which agents are not allowed to remain unmatched (equivalently, we may allow agents to remain single but set their reservations utilities to $-\infty$).

For simplicity, we will use the ITU-logit model once again. Full assignment implies that $\sum_{x\in\mathcal{X}} n_x = \sum_{y\in\mathcal{Y}} m_y$, i.e. there is an equal mass of agents on both sides of the market since all must match. While it is no longer possible to recover the systematic utilities $U_{xy}$ and $V_{xy}$ from the matching patterns, one can show that 
\begin{equation*}
    U_{xy} = \log \mu_{xy} + a_x \text{ and } V_{xy} = \log \mu_{xy} + b_y
\end{equation*}
where $a_x$ and $b_y$ play the role of fixed effects and are given by $a_x=-\log \sum_{y \in \mathcal{Y}} \exp (U_{xy})/n_x$ and $b_y=-\log  \sum_{x \in \mathcal{X}}\exp (V_{xy})/m_y$. The feasibility condition $D_{xy}( U_{xy},V_{xy}) = 0$ from definition \ref{def:equil} yields $\log \mu_{xy} = - D_{xy}(a_x,b_y)$ from which we obtain the matching function $M_{xy}$ which relates the mass of matches of type $xy$ to the fixed effects $a_x$ and $b_y$, that is
\begin{equation*}
    \mu_{xy} = M_{xy}(a_x, b_y)
\end{equation*}
where $M_{xy}(a_x, b_y) = \exp(- D_{xy}(a_x,b_y))$. The analog to the matching function equilibrium with partial assignment in the full assignment case is given below:

\begin{definition}[\cite{chenchoogalichonetal2023}]
\label{def:MFE_full}
 A matching function equilibrium model with full assignment determines the mass $\mu _{xy}$
of $xy\in\mathcal{XY}$ matches by an aggregate matching function
which relates $\mu _{xy}$ to the fixed effects $a_{x}$ and $b_{y}$, that is%
\begin{equation*}
\mu _{xy}=M_{xy}(a_{x},b_{y}),
\end{equation*}%
where the fixed effects $a=\{a_x\}_{x \in \mathcal{X}}$ and $b=\{b_y\}_{y \in \mathcal{Y}}$ satisfy a system of nonlinear accounting
equations
\begin{equation}
\begin{cases}
n_{x}=\sum_{y\in \mathcal{Y}}M_{xy}(a_{x},b_{y}),~\,\forall\,\,  x\in
\mathcal{X},\\
m_{y}=\sum_{x\in \mathcal{X}}M_{xy}(a_{x},b_{y}),~\forall\,\, y\in
\mathcal{Y}.
\end{cases}
\label{eq:MFE_system}
\end{equation}
\end{definition}

\medskip
Note, however, that there is some dependency in system (\ref{eq:MFE_system}), since by assumption $\sum_{x\in\mathcal{X}} n_x = \sum_{y\in\mathcal{Y}} m_y$ and by construction $\sum_{x\in \mathcal{X}}\sum_{y\in \mathcal{Y}}M_{xy}(a_{x},b_{y}) = \sum_{y\in \mathcal{Y}}\sum_{x\in \mathcal{X}}M_{xy}(a_{x},b_{y})$ for any $(a,b)$. Therefore, existence and uniqueness of an equilibrium in matching function equilibrium models with full assignment is all but guaranteed. However, it turns out that under some mild conditions on the matching function $M_{xy}$ (namely, continuity, strict isotonicity and appropriate limiting behavior), an equilibrium exists. In addition, under normalization of one of the $a_{x}$ or $b_{y}$, that equilibrium is unique\footnote{Uniqueness holds under more general normalizations on $(a,b)$, as demonstrated in \textcite{chenchoogalichonetal2023}.}. Therefore, like the partial assignment models, full assignment models are well suited for empirical work, where the estimation methods introduced in section \ref{sec:estimation} can be used.

\subsection{Non-transferable utility}

In this chapter, our definition of non-transferable utility yields the bargaining sets
\begin{equation*}
\mathcal{F}_{ij}=\{ ( u,v) \in \mathbb{R}^{2}:u\leq \alpha
_{ij},v\leq \gamma _{ij}\}.
\end{equation*}%
Unlike traditional NTU models, we allow for utility burning. That is, if $i$ and $j$ are matched, then $u_i \leq \alpha_{ij}$ and $v_i \leq \gamma_{ij}$, when at least one of these inequalities holds as an equality, but not necessarily both. In our model, the stable payoffs are not necessarily the (only) efficient allocation $(\alpha_{ij}, \gamma _{ij})$. A possible justification for allowing utility burning is that if agents on one side of the market are relatively scarce, then the agents on the other side of the market can engage in wasteful competition, which decreases their utility. 

Despite these differences, there is a connection between the NTU stability concept defined in this chapter and NTU stability as it features in the existing literature. In fact, \textcite{galichonhsieh2020} shows that when there is one agent of each type, our notion of NTU equilibrium exactly coincides with traditional NTU stability.

The connection breaks however when there are more than a unit mass of agents per observable type, or when there is unobservable heterogeneity, as in our empirical framework. This explains why the matching function we obtain in the NTU-logit case differs from others in the literature. For example, in \textcite{dagsvik2000} and \textcite{menzel2015}, the matching function, which is based on conventional NTU stability, is given by $\mu _{xy}=\mu _{x0}\mu _{0y}e^{\alpha _{xy}+\gamma _{xy}}$. In contrast, recall that our matching function in the NTU-logit case is $\mu _{xy}=\min \{ \mu _{x0}e^{\alpha _{xy}},\mu _{0y}e^{\gamma
_{xy}}\}$. In the first case, the underlying model is that men and women receive utilities $\alpha _{xy}+\varepsilon _{ij}$ and $\gamma _{xy}+\eta _{ij}$, respectively, where $\varepsilon _{ij}$ and $\eta _{ij}$ are i.i.d.~type~I
extreme value distributed shocks. In the second case, the underlying model is that partners receive utilities
$\alpha _{xy}-w_{xy}^{-}+\varepsilon _{iy}$ and $\gamma
_{xy}-w_{xy}^{+}+\eta _{xj}$ where both $w_{xy}^{-}$ and $w_{xy}^{+}$ are nonnegative and at least one of these terms is equal to $0$ and $\varepsilon _{iy}$ and $\eta _{xj}$ are i.i.d.~type~I extreme value distributed shocks. Note that the NTU model in Dagsvik-Menzel features i.i.d.
preference shocks over individuals, while our model features preference shocks
over types. Therefore, the matching
function from Dagsvik-Menzel is not positive homogenous of degree one: as the mass of individuals per
type increases, the fraction of married individuals increases, whereas it
remains constant in our model.

\subsection{Search-and-matching}
\label{sec:search-and-matching}
The distance function approach allows one to generalize the search-theoretic framework of matching with search frictions à-la Shimer and Smith (\cite{shimer2000assortative}) from the TU to the ITU case, as done in~\textcite{lauermann2020balance}, section 6.2. In that paradigm, at each time pairs $xy$ meet according to some Poisson process and both agents decide whether to match
or not. Both have a reservation utility associated with not
matching which is endogenously determined as the option value of waiting for
a better match. 

More precisely, we consider a continuous-time model where agents discount the future with rate $r$. An unmatched man of
type $x$ experiences random meetings with women of type $y$ as a Poisson
process arrival of intensity $\rho \mu _{0y}$; similarly, an unmatched woman
of type $y$ experiences random meetings with men of type $x$ as a Poisson
process arrival of intensity $\rho \mu _{x0}$. Matches between type $x$ and
type $y$ are exogenously destroyed as a Poisson process of intensity $\delta 
$.

Let $U_{x}$ be the intertemporal value of being type unmatched of type $x$.
We assume that if an agent of type $x$ is matched with a type $y$, then its
intertemporal value is $U_{x}+S_{xy}$. Similarly, we
let $V_{y}$ be the intertemporal value of being type unmatched of type $y$,
and we assume that if an agent of type $y$ is matched with a type $x$, then
its intertemporal value is $V_{y}+S_{yx}$. In a match
between $x$ and $y$, the partners will respectively get $\left(
u_{xy},v_{xy}\right) $ which is on the frontier of $\mathcal{F}_{xy}$ at
each period until a match is dissolved therefore the Bellman equation of
men of type $x$ matched with women of type $y$ is 
\begin{eqnarray*}
r\left( U_{x}+S_{xy}\right)  &=&u_{xy}-\delta S_{xy}, \\
r\left( V_{x}+S_{yx}\right)  &=&v_{xy}-\delta S_{yx}.
\end{eqnarray*}%
and thus $u_{xy}=rU_{x}+\left( r+\delta \right) S_{xy}$ and $%
v_{xy}=rV_{y}+\left( r+\delta \right) S_{yx}$. But as $\left( u_{xy},v_{xy}\right) $ is on the frontier of $\mathcal{F}_{xy}
$, $D_{xy}\left( u_{xy},v_{xy}\right) =0$, and therefore 
\[
D_{xy}\left( rU_{x}+\left( r+\delta \right) S_{xy},rV_{y}+\left( r+\delta
\right) S_{yx}\right) =0.
\]

Given that $\left( rU_{x},rV_{y}\right) $ is the threat point in this bargaining game, if one
assumes Nash bargaining, one has $S_{xy}=S_{yx}$ and therefore 
\[
S_{xy}=-\frac{D_{xy}\left( rU_{x},rV_{y}\right) }{\left( r+\delta \right) }
\]

Therefore, agents match if $S_{xy}\geq 0$, that is if $D_{xy}\left(
rU_{x},rV_{y}\right) \leq 0$. Letting $u_{x}=rU_{x}$ and $v_{y}=rV_{y}$, one
can write the Bellman equation associated with unmatched agents of type $x$
and $y$, respectively%
\begin{eqnarray}
u_{x} &=&\rho \sum_{y\in \mathcal{Y}}\mu _{0y}\max \left\{ 0,-D_{xy}\left(
u_{x},v_{y}\right) \right\}   \label{search-eq-1} \\
v_{y} &=&\rho \sum_{x\in \mathcal{X}}\mu _{x0}\max \left\{ 0,-D_{xy}\left(
u_{x},v_{y}\right) \right\}   \label{search-eq-2}
\end{eqnarray}%
and steady state on the flows of $xy$ matches yield the matching function%
\begin{equation}\label{matching-function-search}
\mu _{xy}=\frac{\rho }{\delta }\mu _{x0}\mu _{0y}1\left\{ D_{xy}\left(
u_{x},v_{y}\right) \leq 0\right\}     
\end{equation}

where the involved unknowns are subject to 
\begin{eqnarray}
n_{x} &=&\mu _{x0}+\sum_{y\in \mathcal{Y}}\frac{\rho }{\delta }\mu _{x0}\mu
_{0y}1\left\{ D_{xy}\left( u_{x},v_{y}\right) \leq 0\right\} 
\label{search-eq-3} \\
m_{y} &=&\mu _{0y}+\sum_{x\in \mathcal{X}}\frac{\rho }{\delta }\mu _{x0}\mu
_{0y}1\left\{ D_{xy}\left( u_{x},v_{y}\right) \leq 0\right\} 
\label{search-eq-4}
\end{eqnarray}
We can therefore extend Shimer and Smith's definition of a \emph{steady-state search equilibrium} from the TU to the ITU case: such equilibrium is an outcome $(\mu_{xy},\mu_{x0},\mu_{0y},u_x,v_y)$ satisfying equations~\eqref{search-eq-1} to~\eqref{search-eq-4} above. A very interesting research avenue consists of incorporating random utility à-la \textcite{choosiow2006} in this type of models of matching with search frictions; an important step in this direction is~\textcite{goussejacquemetrobin2017}.

\section{Related literature}
In the following section, we provide a brief overview of the literature on, or connected to, matching models with imperfectly transferable utility.

\textit{Transferable utility.} At one end of the spectrum are matching models with (perfectly) transferable utility. \textcite{koopmansbeckmann1957} is one of the earliest paper to study the assignment problem and its formulation as a linear programming problem. \textcite{shapleyshubik1971} studies assignment games, with a particular emphasis on the core of such games. \textcite{becker1973} is one of the early application of the assignment problem to marriage markets. Many of the fundamental results from that early literature can be found in \textcite[][chapter 8]{rothsotomayor1990}, \textcite[][chapter 3]{galichon2016} or \textcite[][chapter 7]{browningchiapporiweiss2014}.

\textit{Non-transferable utility.} At the other hand of the spectrum are non-transferable utility models. The classical NTU model is studied in \textcite{galeshapley1962}, along with the deferred-acceptance algorithm that produces a stable matching for such matching markets. NTU models are covered in detail in \textcite{rothsotomayor1990}. \textcite{galichonhsieh2020} examine a NTU framework with utility burning, and therefore provides a comprehensive analysis of the connections between this chapter's treatment of NTU models and the classical NTU model.

\textit{Imperfectly transferable utility.} Several researchers have explored the theory of ITU matching, including \textcite{crawfordknoer1981}, \textcite{kelsocrawford1982}, and \textcite{hatfieldmilgrom2005}. These studies have identified conditions and algorithms for finding competitive equilibria
outcomes in ITU matching markets and have studied the structure of the
sets of equilibria. 

Other notable contributions to ITU matching theory include the work of \textcite{kaneko1982}, \textcite{quinzii1984}, \textcite{alkan1989}, \textcite{alkangale1990}, \textcite{gale1984}, \textcite{demangegale1985}, and \textcite{fleinerjagadeesanjankoetal2019}. The papers provide results on the
existence of equilibria and analyzed properties of the core. \textcite{pycia2012}
features a many-to-one ITU matching framework, and characterizes the sharing rules that lead to pairwise alignment of preferences and existence of equilibria.

More recently, \textcite{noeldekesamuelson2018} examined connections between abstract notions of convexity and matching with ITU, while \textcite{greineckerkah2021} analyzed the notion of pairwise stability in a ITU framework with a continuum of agents. 

Finally, matching models with perfectly, imperfectly and non-transferable utility are surveyed in \textcite{chiappori2017} and \textcite{chadeeeckhoutsmith2017}. The latter also covers search and matching models with non-transferable utility, complementing the ITU search and matching framework introduced in section \ref{sec:search-and-matching}.

\textit{Distance-to-frontier functions.} The approach of describing bargaining sets via distance-to-frontier functions draws its inspiration from the literature on directional distance functions and efficiency in production, see e.g. \textcite{deatonmuellbauer1980}, \textcite{chamberschungfaere1998}, and \textcite{parmeterkumbhakar2014}.

\textit{Existence and uniqueness results.}  The methods introduced in this chapter to establish existence and uniqueness of an equilibrium are based on the notion of gross substitutability, which features prominently in \textcite{kelsocrawford1982} and in the literature on demand inversion problems, e.g. \textcite{berrygandhihaile2013}. Other approaches to establish existence of an equilibrium in the unregularized case include \textcite{scarf1967}, \textcite{kelsocrawford1982}, \textcite{alkan1989}, \textcite{alkangale1990} or \textcite{hatfieldmilgrom2005}, as discussed in section \ref{sec:unregularizedcase}. 

In our setting, uniqueness also hinges on the full support assumption, for similar reasons that uniqueness arises in continuum matching settings, as in \textcite{chiapporimccannnesheim2010} in the TU case or \textcite{azevedoleshno2016} in the NTU case. Other papers have followed a different approach, e.g. \textcite{galichonsalanie2022} rely on convex optimization, albeit in the TU case. 

Our reformulation of the matching model as a demand system is in the same spirit as  \textcite{azevedoleshno2016} in the NTU case without
unobserved heterogeneity. Recently, \textcite{chenchoogalichonetal2023} follow a similar approach (with a supply system) to provide existence and uniqueness results in the full assignment case.

\textit{Assortative matching.} \textcite{legrosnewman2010} provide a sufficient and necessary condition for assortative matching in a strict NTU setting. Our treatment of assortative matching in the ITU case follows \textcite[][chapter 7]{chiappori2017}. \textcite{legrosnewman2007} study assortativeness in ITU models, provides conditions under
which PAM arises, and discuss applications to matching under uncertainty with risk aversion. More recently, \textcite{chiapporireny2016} considered a similar setup with risk sharing, while \textcite{chadeeeckhout2017} extended the setting of \textcite{legrosnewman2007} to the case where agents have different risky endowments. 
Similarly, \textcite{lisunwangetal2016} study assortative matching in two-sided matching markets with risk sharing (but with a focus on the different risks faced by the agents rather than on their different degrees of risk aversion) and provide conditions under which negative assortative matching (or PAM) arises.

\textit{Empirical frameworks.} \textcite{choosiow2006}'s seminal work in the TU case exploits (logit) heterogeneity in tastes for identification purposes, building on the insights of \textcite{mcfadden1974}. They were also the first to realize the analytical convenience of the separability assumption, which decomposes the joint surplus into a systematic component and unobserved heterogeneity components. Many papers have followed in the steps of that paper, e.g. \textcite{fox2010}, \textcite{chiapporiorefficequintanadomeque2012}, \textcite{dupuygalichon2014}, \textcite{chiapporisalanieweiss2017}, \textcite{ciscatoweber2019} and \textcite{galichonsalanie2022}, among others. In the NTU case, several papers have used a similar strategy, e.g. \textcite{dagsvik2000}, \textcite{menzel2015}, \textcite{hitschhortacsuariely2010} or \textcite{agarwal2015}. In the ITU case, the literature is much more sparse. To the best of our knowledge, \textcite{galichonkominersweber2019} was the first paper to provide a tractable empirical framework for matching with ITU.

Finally, the ITU-logit model presented in this chapter is reminiscent of the demography literature on matching functions, e.g. \textcite{schoen1981}. In economics, matching functions have been studied in detail in \textcite{siow2008}, \textcite{mourifie2019} and \textcite{mourifiesiow2021}. \textcite{chenchoogalichonetal2023b} define and study the properties of matching function equilibria, which consists of a behaviorally coherent matching function and accounting constraints, and encompasses many models in this literature.

\textit{Estimation and computation.} Estimation of the ITU-logit model by maximum likelihood is described in \textcite{galichonkominersweber2019}. \textcite{chenchoogalichonetal2023} and \textcite{chenchoogalichonetal2023b} discuss the estimation of matching function equibria models (with full assignment and partial assignment, respectively) by maximum likelihood. Algorithm \ref{algo:ippf} used in those papers is reminiscent of the iterative projective fitting procedure algorithms used in other literatures under different names, see \textcite{idel2016} for a review. In the TU-logit case, algorithm \ref{algo:ippf} simplifies greatly, as shown in \textcite{galichonsalanie2022}. 


\textit{Marriage markets.} The collective approach to household decision making, introduced in \textcite{chiappori1988}, \textcite{appsrees1988} and \textcite{chiappori1992}, assumes that decision makers in household bargain and take Pareto-efficient decisions \parencite[see][for a discussion on the efficiency assumption]{delbocaflinn2012}. This approach has spurned a vast theoretical and empirical literature and collective models have quickly become a cornerstone of family economics \parencite[see][for a review]{browningchiapporiweiss2014}. However, in these models, both household formation and the allocation of bargaining power are taken as given. Therefore, a growing literature has attempted to embed collective models into matching models of the marriage market in order to endogenize family formation and the allocation of bargaining power. Most of the early papers reconciling the collective and matching approaches have done so in the TU case, e.g. \textcite{iyigunwalsh2007}, \textcite{chiapporiiyigunweiss2009}, \textcite{chooseitz2013}, \textcite{chiapporidiasmeghir2018} \parencite[see also][chapters 8 and 9 for a review]{browningchiapporiweiss2014}. In the ITU case, \textcite{chiappori2012} provides an illustrative example on how to embed a collective model into a matching framework \parencite[see also][chapter 7]{chiappori2017}. \textcite{galichonkominersweber2019} provide an empirical matching framework with ITU and features several examples inspired by the collective approach. \textcite{weber2022} discusses further the connection between collective models and ITU models, and characterizes ``proper'' collective models that can be embedded in the empirical framework described in this chapter. \textcite{reynoso2023} extends the framework to incorporate a life cycle model and provides an application.

While the framework introduced in this chapter is frictionless, other papers in the search and matching literature have examined the connection between collective models and marriage markets. For example,~\textcite{goussejacquemetrobin2017} focuses on the TU case while \textcite{colesfrancesconi2019} as well~\textcite{lauermann2020balance} considers the ITU case and provides an application with NTU.

Finally, other papers in the literature have followed a completely different approach, e.g. \textcite{cherchyedemuynckderocketal2017} derive Afriat-style inequalities from stability conditions on the marriage market to obtain bounds on intra-household sharing rules, which describe how resources are allocated within households.

\textit{Labour markets.} \textcite{dupuygalichonjaffeetal2020} study an ITU matching problem with linear taxes which is reformulated as a TU matching problem, and provide
comparative statics results as well as an application to the college-coach football labour market. \textcite{corblet2023} extends the \textcite{choosiow2006} framework to one-to-many matching and provides an application to the Portuguese labour market.

\clearpage
\newpage
\printbibliography

@Article{petrongolopissarides2001,
  author    = {Petrongolo, Barbara and Pissarides, Christopher A},
  journal   = {Journal of Economic literature},
  title     = {Looking into the black box: A survey of the matching function},
  year      = {2001},
  number    = {2},
  pages     = {390--431},
  volume    = {39},
  publisher = {American Economic Association},
}

@Article{galichonkominersweber2019,
  author   = {Galichon, Alfred and Kominers, Scott Duke and Weber, Simon},
  journal  = {Journal of Political Economy},
  title    = {Costly Concessions: An Empirical Framework for Matching with Imperfectly Transferable Utility},
  year     = {2019},
  number   = {6},
  pages    = {2875-2925},
  volume   = {127},
}

@Article{becker1973,
  author   = {Becker, Gary S.},
  journal  = {Journal of Political Economy},
  title    = {A Theory of Marriage: Part {I}},
  year     = {1973},
  number   = {4},
  pages    = {813-846},
  volume   = {81},
}

@Article{koopmansbeckmann1957,
  author    = {Tjalling C. Koopmans and Martin Beckmann},
  journal   = {Econometrica},
  title     = {Assignment Problems and the Location of Economic Activities},
  year      = {1957},
  issn      = {00129682, 14680262},
  number    = {1},
  pages     = {53--76},
  volume    = {25},
  publisher = {[Wiley, Econometric Society]},
  urldate   = {2023-11-30},
}

@Article{shapleyshubik1971,
  author    = {Shapley, L. S. and Shubik, M.},
  journal   = {International Journal of Game Theory},
  title     = {The assignment game {I}: The core},
  year      = {1971},
  issn      = {1432-1270},
  month     = dec,
  number    = {1},
  pages     = {111–130},
  volume    = {1},
  publisher = {Springer Science and Business Media LLC},
}

@Book{rothsotomayor1990,
  author     = {Roth, Alvin E. and Sotomayor, Marilda A. Oliveira},
  publisher  = {Cambridge University Press},
  title      = {Two-Sided Matching: A Study in Game-Theoretic Modeling and Analysis},
  year       = {1990},
  series     = {Econometric Society Monographs},
  collection = {Econometric Society Monographs},
  place      = {Cambridge},
}

@Book{chiappori2017,
  author    = {Chiappori, Pierre-André},
  publisher = {Princeton University Press},
  title     = {Matching with Transfers: The Economics of Love and Marriage},
  year      = {2017},
  isbn      = {9781400885732},
  month     = may,
}

@Book{galichon2016,
  author    = {Alfred Galichon},
  publisher = {Princeton University Press},
  title     = {Optimal Transport Methods in Economics},
  year      = {2016},
  urldate   = {2023-11-30},
}

@Book{browningchiapporiweiss2014,
  author     = {Browning, Martin and Chiappori, Pierre-André and Weiss, Yoram},
  publisher  = {Cambridge University Press},
  title      = {Economics of the Family},
  year       = {2014},
  series     = {Cambridge Surveys of Economic Literature},
  collection = {Cambridge Surveys of Economic Literature},
  place      = {Cambridge},
}

@Article{galeshapley1962,
  author    = {D. Gale and L. S. Shapley},
  journal   = {The American Mathematical Monthly},
  title     = {College Admissions and the Stability of Marriage},
  year      = {1962},
  issn      = {00029890, 19300972},
  number    = {1},
  pages     = {9--15},
  volume    = {69},
  publisher = {Mathematical Association of America},
  urldate   = {2023-11-30},
}

@article{lauermann2020balance,
  title={The Balance Condition in Search-and-Matching Models},
  author={Lauermann, Stephan and N{\"o}ldeke, Georg and Tr{\"o}ger, Thomas},
  journal={Econometrica},
  volume={88},
  number={2},
  pages={595--618},
  year={2020},
  publisher={Wiley Online Library}
}

@Article{galichonhsieh2020,
  author  = {Galichon, Alfred and Hsieh, Yu-Wei and Sylvestre, Maxime},
  journal = {working paper},
  title   = {Aggregate Stable Matching with Money Burning},
  year    = {2020},
}

@Article{alkan1989,
  author   = {Ahmet Alkan},
  journal  = {European Journal of Political Economy},
  title    = {Existence and computation of matching equilibria},
  year     = {1989},
  issn     = {0176-2680},
  number   = {2},
  pages    = {285-296},
  volume   = {5},
}

@Article{alkangale1990,
  author   = {Ahmet Alkan and David Gale},
  journal  = {Games and Economic Behavior},
  title    = {The core of the matching game},
  year     = {1990},
  issn     = {0899-8256},
  number   = {3},
  pages    = {203-212},
  volume   = {2},
}

@article{shimer2000assortative,
  title={Assortative matching and search},
  author={Shimer, Robert and Smith, Lones},
  journal={Econometrica},
  volume={68},
  number={2},
  pages={343--369},
  year={2000},
  publisher={Wiley Online Library}
}

@Article{crawfordknoer1981,
  author    = {Vincent P. Crawford and Elsie Marie Knoer},
  journal   = {Econometrica},
  title     = {Job Matching with Heterogeneous Firms and Workers},
  year      = {1981},
  issn      = {00129682, 14680262},
  number    = {2},
  pages     = {437--450},
  volume    = {49},
  publisher = {[Wiley, Econometric Society]},
  urldate   = {2023-12-01},
}

@Article{demangegale1985,
  author    = {Gabrielle Demange and David Gale},
  journal   = {Econometrica},
  title     = {The Strategy Structure of Two-Sided Matching Markets},
  year      = {1985},
  issn      = {00129682, 14680262},
  number    = {4},
  pages     = {873--888},
  volume    = {53},
  publisher = {[Wiley, Econometric Society]},
  urldate   = {2023-12-01},
}

@Article{fleinerjagadeesanjankoetal2019,
  author   = {Fleiner, Tamás and Jagadeesan, Ravi and Jankó, Zsuzsanna and Teytelboym, Alexander},
  journal  = {Econometrica},
  title    = {Trading Networks With Frictions},
  year     = {2019},
  number   = {5},
  pages    = {1633-1661},
  volume   = {87},
}

@Article{gale1984,
  author    = {Gale, D.},
  journal   = {International Journal of Game Theory},
  title     = {Equilibrium in a discrete exchange economy with money},
  year      = {1984},
  issn      = {1432-1270},
  month     = mar,
  number    = {1},
  pages     = {61–64},
  volume    = {13},
  publisher = {Springer Science and Business Media LLC},
}

@Article{hatfieldmilgrom2005,
  author  = {Hatfield, John William and Milgrom, Paul R.},
  journal = {American Economic Review},
  title   = {Matching with Contracts},
  year    = {2005},
  month   = {September},
  number  = {4},
  pages   = {913-935},
  volume  = {95},
}

@Article{kaneko1982,
  author   = {Mamoru Kaneko},
  journal  = {Journal of Mathematical Economics},
  title    = {The central assignment game and the assignment markets},
  year     = {1982},
  issn     = {0304-4068},
  number   = {2},
  pages    = {205-232},
  volume   = {10},
}

@Article{kelsocrawford1982,
  author    = {Alexander S. Kelso and Vincent P. Crawford},
  journal   = {Econometrica},
  title     = {Job Matching, Coalition Formation, and Gross Substitutes},
  year      = {1982},
  issn      = {00129682, 14680262},
  number    = {6},
  pages     = {1483--1504},
  volume    = {50},
  publisher = {[Wiley, Econometric Society]},
  urldate   = {2023-12-01},
}

@Article{pycia2012,
  author   = {Pycia, Marek},
  journal  = {Econometrica},
  title    = {Stability and Preference Alignment in Matching and Coalition Formation},
  year     = {2012},
  number   = {1},
  pages    = {323-362},
  volume   = {80},
}

@Article{quinzii1984,
  author    = {Quinzii, M.},
  journal   = {International Journal of Game Theory},
  title     = {Core and competitive equilibria with indivisibilities},
  year      = {1984},
  issn      = {1432-1270},
  month     = mar,
  number    = {1},
  pages     = {41–60},
  volume    = {13},
  publisher = {Springer Science and Business Media LLC},
}

@Article{greineckerkah2021,
  author   = {Greinecker, Michael and Kah, Christopher},
  journal  = {Econometrica},
  title    = {Pairwise Stable Matching in Large Economies},
  year     = {2021},
  number   = {6},
  pages    = {2929-2974},
  volume   = {89},
}

@Article{noeldekesamuelson2018,
  author   = {Nöldeke, Georg and Samuelson, Larry},
  journal  = {Econometrica},
  title    = {The Implementation Duality},
  year     = {2018},
  number   = {4},
  pages    = {1283-1324},
  volume   = {86},
}

@Book{deatonmuellbauer1980,
  author    = {Deaton, Angus and Muellbauer, John},
  publisher = {Cambridge University Press},
  title     = {Economics and Consumer Behavior},
  year      = {1980},
  place     = {Cambridge},
}

@Article{chamberschungfaere1998,
  author    = {Chambers, R. G. and Chung, Y. and Färe, R.},
  journal   = {Journal of Optimization Theory and Applications},
  title     = {Profit, Directional Distance Functions, and {Nerlovian} Efficiency},
  year      = {1998},
  issn      = {1573-2878},
  month     = aug,
  number    = {2},
  pages     = {351–364},
  volume    = {98},
  publisher = {Springer Science and Business Media LLC},
}

@Article{parmeterkumbhakar2014,
  author  = {Christopher F. Parmeter and Subal C. Kumbhakar},
  journal = {Foundations and Trends® in Econometrics},
  title   = {Efficiency Analysis: A Primer on Recent Advances},
  year    = {2014},
  issn    = {1551-3076},
  number  = {3–4},
  pages   = {191-385},
  volume  = {7},
}

@Article{azevedoleshno2016,
  author   = {Azevedo, Eduardo M. and Leshno, Jacob D.},
  journal  = {Journal of Political Economy},
  title    = {A Supply and Demand Framework for Two-Sided Matching Markets},
  year     = {2016},
  number   = {5},
  pages    = {1235-1268},
  volume   = {124},
}

@Article{chiapporimccannnesheim2010,
  author    = {Pierre-André Chiappori and Robert J. McCann and Lars P. Nesheim},
  journal   = {Economic Theory},
  title     = {Hedonic Price Equilibria, Stable Matching, and Optimal Transport: Equivalence, Topology, and Uniqueness},
  year      = {2010},
  issn      = {09382259, 14320479},
  number    = {2},
  pages     = {317--354},
  volume    = {42},
  publisher = {Springer},
  urldate   = {2023-12-04},
}

@Article{galichonsalanie2022,
  author   = {Galichon, Alfred and Salanié, Bernard},
  journal  = {The Review of Economic Studies},
  title    = {{Cupid’s Invisible Hand: Social Surplus and Identification in Matching Models}},
  year     = {2022},
  issn     = {0034-6527},
  month    = {12},
  number   = {5},
  pages    = {2600-2629},
  volume   = {89},
  eprint   = {https://academic.oup.com/restud/article-pdf/89/5/2600/45764005/rdab090.pdf},
}

@Article{berrygandhihaile2013,
  author   = {Berry, Steven and Gandhi, Amit and Haile, Philip},
  journal  = {Econometrica},
  title    = {Connected Substitutes and Invertibility of Demand},
  year     = {2013},
  number   = {5},
  pages    = {2087-2111},
  volume   = {81},
}

@Article{legrosnewman2007,
  author   = {Legros, Patrick and Newman, Andrew F.},
  journal  = {Econometrica},
  title    = {Beauty Is a Beast, Frog Is a Prince: Assortative Matching with Nontransferabilities},
  year     = {2007},
  number   = {4},
  pages    = {1073-1102},
  volume   = {75},
}

@Article{chiapporireny2016,
  author   = {Chiappori, Pierre-André and Reny, Philip J.},
  journal  = {Theoretical Economics},
  title    = {Matching to share risk},
  year     = {2016},
  number   = {1},
  pages    = {227-251},
  volume   = {11},
}

@Article{chadeeeckhout2017,
  author  = {Hector Chade and Jan Eeckhout},
  journal = {working paper},
  title   = {Stochastic Sorting},
  year    = {2017},
}

@Article{chiapporiorefficequintanadomeque2012,
  author   = {Chiappori, Pierre-Andr\'{e} and Oreffice, Sonia and Quintana-Domeque, Climent},
  journal  = {Journal of Political Economy},
  title    = {Fatter Attraction: Anthropometric and Socioeconomic Matching on the Marriage Market},
  year     = {2012},
  number   = {4},
  pages    = {659-695},
  volume   = {120},
}

@Article{choosiow2006,
  author   = {Choo, Eugene and Siow, Aloysius},
  journal  = {Journal of Political Economy},
  title    = {Who Marries Whom and Why},
  year     = {2006},
  number   = {1},
  pages    = {175-201},
  volume   = {114},
}

@Article{ciscatoweber2019,
  author    = {Ciscato, Edoardo and Weber, Simon},
  journal   = {Journal of Population Economics},
  title     = {The role of evolving marital preferences in growing income inequality},
  year      = {2019},
  issn      = {1432-1475},
  month     = jul,
  number    = {1},
  pages     = {307–347},
  volume    = {33},
  publisher = {Springer Science and Business Media LLC},
}

@Article{dupuygalichon2014,
  author   = {Dupuy, Arnaud and Galichon, Alfred},
  journal  = {Journal of Political Economy},
  title    = {Personality Traits and the Marriage Market},
  year     = {2014},
  number   = {6},
  pages    = {1271-1319},
  volume   = {122},
}

@Article{chiapporisalanieweiss2017,
  author  = {Chiappori, Pierre-André and Salanié, Bernard and Weiss, Yoram},
  journal = {American Economic Review},
  title   = {Partner Choice, Investment in Children, and the Marital College Premium},
  year    = {2017},
  month   = {August},
  number  = {8},
  pages   = {2109-67},
  volume  = {107},
}

@InCollection{mcfadden1974,
  author    = {Daniel McFadden},
  booktitle = {Frontiers in Econometrics},
  publisher = {New York: Academic Press},
  title     = {Conditional Logit Analysis of Qualitative Choice Behavior},
  year      = {1974},
  editor    = {Paul Zarembka},
  pages     = {105–42},
}

@Article{dagsvik2000,
  author   = {Dagsvik, John K.},
  journal  = {International Economic Review},
  title    = {Aggregation in Matching Markets},
  year     = {2000},
  number   = {1},
  pages    = {27-58},
  volume   = {41},
}

@Article{menzel2015,
  author   = {Menzel, Konrad},
  journal  = {Econometrica},
  title    = {Large Matching Markets as Two-Sided Demand Systems},
  year     = {2015},
  number   = {3},
  pages    = {897-941},
  volume   = {83},
}

@Article{agarwal2015,
  author  = {Agarwal, Nikhil},
  journal = {American Economic Review},
  title   = {An Empirical Model of the Medical Match},
  year    = {2015},
  month   = {July},
  number  = {7},
  pages   = {1939-78},
  volume  = {105},
}

@incollection{noeldeke_samuelson_handbook,
  author       = {Nöldeke, Georg and Samuelson, Larry},
  title        = {Investment and Competitive Matching},
  booktitle    = {Handbook of the Economics of Matching},
  editor       = {Y.-K. Che and P.-A. Chiappori and B. Salanié},
  year         = {2024},
  publisher    = {Elsevier},
  address      = {Amsterdam},
}

@incollection{salanie_handbook,
  author       = {Salanié, Bernard},
  title        = {Matching with Transfers: Applications},
  booktitle    = {Handbook of the Economics of Matching},
  editor       = {Y.-K. Che and P.-A. Chiappori and B. Salanié},
  year         = {2024},
  publisher    = {Elsevier},
  address      = {Amsterdam},
}

@incollection{chiappori_low_handbook,
  author       = {Chiappori, Pierre-André, and Low, Corinne},
  title        = {Matching with Transfers: Theory},
  booktitle    = {Handbook of the Economics of Matching},
  editor       = {Y.-K. Che and P.-A. Chiappori and B. Salanié},
  year         = {2024},
  publisher    = {Elsevier},
  address      = {Amsterdam},
}

@incollection{hatfield_kominers_handbook,
  author       = {Hatfield, John, and Kominers, Scott},
  title        = {Matching with Transfers: Theory},
  booktitle    = {Handbook of the Economics of Matching},
  editor       = {Y.-K. Che and P.-A. Chiappori and B. Salanié},
  year         = {2024},
  publisher    = {Elsevier},
  address      = {Amsterdam},
}

@Article{hitschhortacsuariely2010,
  author  = {Hitsch, Gunter J. and Hortaçsu, Ali and Ariely, Dan},
  journal = {American Economic Review},
  title   = {Matching and Sorting in Online Dating},
  year    = {2010},
  month   = {March},
  number  = {1},
  pages   = {130-63},
  volume  = {100},
}

@Article{cherchyedemuynckderocketal2017,
  author  = {Cherchye, Laurens and Demuynck, Thomas and De Rock, Bram and Vermeulen, Frederic},
  journal = {American Economic Review},
  title   = {Household Consumption When the Marriage Is Stable},
  year    = {2017},
  month   = {June},
  number  = {6},
  pages   = {1507-34},
  volume  = {107},
}

@Article{fox2010,
  author   = {Fox, Jeremy T.},
  journal  = {Quantitative Economics},
  title    = {Identification in matching games},
  year     = {2010},
  number   = {2},
  pages    = {203-254},
  volume   = {1},
}

@Article{appsrees1988,
  author   = {Patricia F. Apps and Ray Rees},
  journal  = {Journal of Public Economics},
  title    = {Taxation and the household},
  year     = {1988},
  issn     = {0047-2727},
  number   = {3},
  pages    = {355-369},
  volume   = {35},
}

@Article{chiappori1988,
  author    = {Pierre-André Chiappori},
  journal   = {Econometrica},
  title     = {Rational Household Labor Supply},
  year      = {1988},
  issn      = {00129682, 14680262},
  number    = {1},
  pages     = {63--90},
  volume    = {56},
  publisher = {[Wiley, Econometric Society]},
  urldate   = {2023-12-05},
}

@Article{chiappori1992,
  author   = {Chiappori, Pierre-Andr\'{e}},
  journal  = {Journal of Political Economy},
  title    = {Collective Labor Supply and Welfare},
  year     = {1992},
  number   = {3},
  pages    = {437-467},
  volume   = {100},
}

@Article{chiappori2012,
  author    = {Pierre-André Chiappori},
  journal   = {Revue économique},
  title     = {Modèles d'appariement en économie: Quelques avancées récentes},
  year      = {2012},
  issn      = {00352764, 19506694},
  number    = {3},
  pages     = {437--452},
  volume    = {63},
  publisher = {Sciences Po University Press},
  urldate   = {2023-12-05},
}

@Article{chiapporiiyigunweiss2009,
  author  = {Chiappori, Pierre-André and Iyigun, Murat and Weiss, Yoram},
  journal = {American Economic Review},
  title   = {Investment in Schooling and the Marriage Market},
  year    = {2009},
  month   = {December},
  number  = {5},
  pages   = {1689-1713},
  volume  = {99},
}

@Article{delbocaflinn2012,
  author   = {Daniela {Del Boca} and Christopher Flinn},
  journal  = {Journal of Econometrics},
  title    = {Endogenous household interaction},
  year     = {2012},
  issn     = {0304-4076},
  note     = {Annals Issue on ``Identification and Decisions'', in Honor of Chuck Manski's 60th Birthday},
  number   = {1},
  pages    = {49-65},
  volume   = {166},
}

@Article{mourifie2019,
  author    = {Mourifié, Ismael},
  journal   = {Economic Theory},
  title     = {A marriage matching function with flexible spillover and substitution patterns},
  year      = {2019},
  issn      = {1432-0479},
  month     = sep,
  number    = {2},
  pages     = {421–461},
  volume    = {67},
  publisher = {Springer Science and Business Media LLC},
}

@Article{mourifiesiow2021,
  author   = {Mourifi\'{e}, Ismael and Siow, Aloysius},
  journal  = {Journal of Labor Economics},
  title    = {The {Cobb-Douglas} Marriage Matching Function: Marriage Matching with Peer and Scale Effects},
  year     = {2021},
  number   = {S1},
  pages    = {S239-S274},
  volume   = {39},
}

@Article{schoen1981,
  author    = {Robert Schoen},
  journal   = {Demography},
  title     = {The Harmonic Mean as the Basis of a Realistic Two-Sex Marriage Model},
  year      = {1981},
  issn      = {00703370, 15337790},
  number    = {2},
  pages     = {201--216},
  volume    = {18},
  publisher = {Springer},
  urldate   = {2023-12-05},
}

@Article{siow2008,
  author   = {Siow, Aloysius},
  journal  = {Canadian Journal of Economics/Revue canadienne d'économique},
  title    = {How does the marriage market clear? An empirical framework},
  year     = {2008},
  number   = {4},
  pages    = {1121-1155},
  volume   = {41},
}

@Article{weber2022,
  author  = {Weber, Simon},
  journal = {working paper},
  title   = {Collective models and the marriage market},
  year    = {2022},
}

@Article{dupuygalichonjaffeetal2020,
  author   = {Dupuy, Arnaud and Galichon, Alfred and Jaffe, Sonia and Kominers, Scott Duke},
  journal  = {International Economic Review},
  title    = {Taxation in Matching Markets},
  year     = {2020},
  number   = {4},
  pages    = {1591-1634},
  volume   = {61},
}

@Article{iyigunwalsh2007,
  author    = {Murat Iyigun and Randall P. Walsh},
  journal   = {The Review of Economic Studies},
  title     = {Building the Family Nest: Premarital Investments, Marriage Markets, and Spousal Allocations},
  year      = {2007},
  issn      = {00346527, 1467937X},
  number    = {2},
  pages     = {507--535},
  volume    = {74},
  publisher = {[Oxford University Press, Review of Economic Studies, Ltd.]},
  urldate   = {2023-12-05},
}

@Article{chiapporidiasmeghir2018,
  author   = {Chiappori, Pierre-Andr\'{e} and Dias, Monica Costa and Meghir, Costas},
  journal  = {Journal of Political Economy},
  title    = {The Marriage Market, Labor Supply, and Education Choice},
  year     = {2018},
  number   = {S1},
  pages    = {S26-S72},
  volume   = {126},
}

@Article{colesfrancesconi2019,
  author   = {Coles, Melvyn G. and Francesconi, Marco},
  journal  = {Journal of Political Economy},
  title    = {Equilibrium Search with Multiple Attributes and the Impact of Equal Opportunities for Women},
  year     = {2019},
  number   = {1},
  pages    = {138-162},
  volume   = {127},
}

@Article{goussejacquemetrobin2017,
  author   = {Marion Goussé and Nicolas Jacquemet and Jean-Marc Robin},
  journal  = {Labour Economics},
  title    = {Household labour supply and the marriage market in the {UK}, 1991-2008},
  year     = {2017},
  issn     = {0927-5371},
  pages    = {131-149},
  volume   = {46},
}

@InCollection{chooseitz2013,
  author    = {Choo, Eugene and Seitz, Shannon},
  publisher = {Emerald Group Publishing Limited},
  title     = {The Collective Marriage Matching Model: Identification, Estimation, and Testing},
  year      = {2013},
  month     = jan,
  pages     = {291--336},
  series    = {Advances in Econometrics},
  volume    = {31},
  issn      = {978-1-78350-052-9},
  journal   = {Structural Econometric Models},
}

@Article{reynoso2023,
  author  = {Ana Reynoso},
  journal = {working paper},
  title   = {The impact of divorce laws on the equilibriumin the marriage market},
  year    = {2023},
}

@Article{ciscatogalichongousse2020,
  author   = {Ciscato, Edoardo and Galichon, Alfred and Gouss\'{e}, Marion},
  journal  = {Journal of Political Economy},
  title    = {Like Attract Like? A Structural Comparison of Homogamy across Same-Sex and Different-Sex Households},
  year     = {2020},
  number   = {2},
  pages    = {740-781},
  volume   = {128},
}

@Book{becker1993,
  author    = {Becker, Gary S.},
  publisher = {Harvard University Press},
  title     = {A Treatise on the Family: Enlarged Edition},
  year      = {1993},
  isbn      = {9780674906983},
  month     = oct,
}

@Article{deckerliebmccannetal2013,
  author   = {Colin Decker and Elliott H. Lieb and Robert J. McCann and Benjamin K. Stephens},
  journal  = {Journal of Economic Theory},
  title    = {Unique equilibria and substitution effects in a stochastic model of the marriage market},
  year     = {2013},
  issn     = {0022-0531},
  number   = {2},
  pages    = {778-792},
  volume   = {148},
}

@Article{chenchoogalichonetal2023,
  author        = {Liang Chen and Eugene Choo and Alfred Galichon and Simon Weber},
  journal       = {working paper},
  title         = {Existence of a Competitive Equilibrium with Substitutes, with Applications to Matching and Discrete Choice Models},
  year          = {2023},
  archiveprefix = {arXiv},
  eprint        = {2309.11416},
  primaryclass  = {econ.GN},
}

@Article{chenchoogalichonetal2023b,
  author        = {Liang Chen and Eugene Choo and Alfred Galichon and Simon Weber},
  journal       = {working paper},
  title         = {Matching Function Equilibria with Partial Assignment: Existence, Uniqueness and Estimation},
  year          = {2023},
  archiveprefix = {arXiv},
  eprint        = {2102.02071},
  primaryclass  = {econ.GN},
}

@Book{rockafellar1970,
  author    = {Rockafellar, Ralph Tyrell},
  publisher = {Princeton University Press},
  title     = {Convex Analysis},
  year      = {1970},
  isbn      = {9781400873173},
  month     = dec,
}

@Article{scarf1967,
  author    = {Herbert E. Scarf},
  journal   = {Econometrica},
  title     = {The Core of an {N} Person Game},
  year      = {1967},
  issn      = {00129682, 14680262},
  number    = {1},
  pages     = {50--69},
  volume    = {35},
  publisher = {[Wiley, Econometric Society]},
  urldate   = {2023-12-22},
}

@Article{lemkehowson1964,
  author    = {C. E. Lemke and J. T. Howson},
  journal   = {Journal of the Society for Industrial and Applied Mathematics},
  title     = {Equilibrium Points of Bimatrix Games},
  year      = {1964},
  issn      = {03684245},
  number    = {2},
  pages     = {413--423},
  volume    = {12},
  publisher = {Society for Industrial and Applied Mathematics},
  urldate   = {2023-12-22},
}

@InProceedings{galichonkominersweber2015,
  author    = {Galichon, Alfred and Kominers, Scott Duke and Weber, Simon},
  booktitle = {Geometric Science of Information},
  title     = {The Nonlinear {Bernstein-Schr{\"o}dinger} Equation in Economics},
  year      = {2015},
  address   = {Cham},
  editor    = {Nielsen, Frank and Barbaresco, Fr{\'e}d{\'e}ric},
  pages     = {51--59},
  publisher = {Springer International Publishing},
  isbn      = {978-3-319-25040-3},
}

@Article{idel2016,
  author        = {Martin Idel},
  title         = {A review of matrix scaling and {Sinkhorn}'s normal form for matrices and positive maps},
  year          = {2016},
  archiveprefix = {arXiv},
  eprint        = {1609.06349},
  primaryclass  = {math.RA},
}

@Article{corblet2023,
  author   = {Corblet, Pauline},
  journal  = {working paper},
  title    = {Education Expansion, Sorting, and the Decreasing Education Wage Premium},
  year     = {2023}
}

@article{chadeeeckhoutsmith2017,
Author = {Chade, Hector and Eeckhout, Jan and Smith, Lones},
Title = {Sorting through Search and Matching Models in Economics},
Journal = {Journal of Economic Literature},
Volume = {55},
Number = {2},
Year = {2017},
Month = {June},
Pages = {493–544},
}

@Article{legrosnewman2010,
  author   = {Patrick Legros and Andrew Newman},
  journal  = {Economics Letters},
  title    = {Co-ranking mates: Assortative matching in marriage markets},
  year     = {2010},
  issn     = {0165-1765},
  number   = {3},
  pages    = {177-179},
  volume   = {106},
}

@Article{lisunwangetal2016,
  author   = {Sanxi Li and Hailin Sun and Tong Wang and Jun Yu},
  journal  = {Journal of Economic Theory},
  title    = {Assortative matching and risk sharing},
  year     = {2016},
  issn     = {0022-0531},
  pages    = {248-275},
  volume   = {163},
}

@Article{prescotttownsend1984,
  author    = {Edward C. Prescott and Robert M. Townsend},
  journal   = {International Economic Review},
  title     = {General Competitive Analysis in an Economy with Private Information},
  year      = {1984},
  issn      = {00206598, 14682354},
  number    = {1},
  pages     = {1--20},
  volume    = {25},
  publisher = {[Economics Department of the University of Pennsylvania, Wiley, Institute of Social and Economic Research, Osaka University]},
  urldate   = {2024-07-02},
}

@Article{williams1977,
  author   = {H C W L Williams},
  journal  = {Environment and Planning A: Economy and Space},
  title    = {On the Formation of Travel Demand Models and Economic Evaluation Measures of User Benefit},
  year     = {1977},
  number   = {3},
  pages    = {285-344},
  volume   = {9},
}

@InCollection{dalyzachary1978,
  author    = {Daly, A. and Zachary, S.},
  booktitle = {Identifying and Measuring the Determinants of Mode Choice},
  publisher = {London: Teakfields},
  title     = {Improved Multiple Choice Models},
  year      = {1978},
  editor    = {Henscher, D. and Dalvi, Q.},
}

\end{document}